\documentclass[paper]{JHEP3} 

\JHEPspecialurl{http://jhep.sissa.it/JOURNAL/JHEP3.tar.gz}

\usepackage{epsfig,multicol}

\newcommand\fverb{\setbox\pippobox=\hbox\bgroup\verb}
\newcommand\fverbdo{\egroup\medskip\noindent%
			\fbox{\unhbox\pippobox}\ }
\newcommand\fverbit{\egroup\item[\fbox{\unhbox\pippobox}]}
\newcommand {\beq}{\begin{equation}}
\newcommand {\eeq}{\end{equation}}
\newcommand {\bea}{\begin{eqnarray}}
\newcommand {\eea}{\end{eqnarray}}
\newcommand {\nn}{\nonumber}
\newcommand {\tr}{{\rm tr\,}}

\newcommand {\dd}{\mbox{d}}

\newcommand {\del}{\partial}
\newcommand {\diag}{\mbox{diag}}

\newcommand {\bA}{\tt A}
\newcommand {\bB}{\tt B}
\newcommand {\bC}{\tt C}
\newcommand {\limit}{\rightarrow}

\newbox\pippobox

\title{A Lattice Formulation of Super Yang-Mills Theories with 
Exact Supersymmetry}

\author{Fumihiko Sugino\\
School of Physics \& BK-21 Physics Division\\
Seoul National University, 
Seoul 151-747, Korea\\
E-mail: \email{sugino@phya.snu.ac.kr}}

\preprint{SNUST-031101 \\\heplat{0311021}}	

\abstract{We construct super Yang-Mills theories 
with extended supersymmetry on hypercubic lattices of various dimensions 
keeping one or two supercharges exactly. 
Gauge fields are represented by ordinary unitary link variables, 
and the exact supercharges are nilpotent up to gauge transformations.  
Among the models, we show that the 
desired continuum theories are obtained without any fine tuning of parameters 
for the cases ${\cal N}=2, 4, 8$ in two-dimensions. 
}


\begin{document} 


\section{Introduction}

Nonperturbative aspects 
in supersymmetric gauge theory have been vigorously investigated, 
in particular since finding the exact form of the low energy effective action 
in four-dimensional ${\cal N}=2$ super Yang-Mills theory \cite{seiberg-witten} 
and the AdS/CFT duality between gauge theory and gravity (string theory) 
\cite{maldacena}. 
For some operators such as the BPS saturated objects or the chiral operators, 
which are protected from the quantum corrections 
by certain algebraic properties, 
we can know exact results for their spectra or 
the correlation functions. 
The approach to the nonperturbative physics is based on the property of 
duality. 

A conventional but more universal approach is lattice formulation, 
which enables numerical analysis for any 
observables not restricted to special operators with 
the BPS saturated or chiral properties. 
However, there has been difficulty on the lattice approach to supersymmetry, 
because of lack of infinitesimal translational invariance on the lattice
and breakdown of the Leibniz rule \cite{fujikawa, elitzur}. 
In spite of the difficulty, it is possible to 
construct lattice models, which do not have manifest supersymmetry but 
flow to the desired supersymmetric theories in the continuum limit. 
One of the examples is ${\cal N}=1$ super Yang-Mills (SYM) theory in 
four-dimensions whose field contents are gauge bosons and gauginos. 
Since in the theory 
the only relevant supersymmetry breaking operator is the gaugino mass, 
one can arrive at the supersymmetric 
continuum theory if the radiative corrections are 
not allowed to induce the relevant operator by symmetries realized  
in the lattice theory. Making use of domain wall or overlap fermions 
keeps discrete chiral symmetry on lattice, which is the symmetry excluding 
the fermion mass \cite{nishimura}.    

Supersymmetric theories with extended supersymmetry have some 
supercharges, which are not related to the infinitesimal translations 
and can be seen as fermionic internal symmetries. It is possible to realize 
a part of such supercharges as exact symmetry on lattice, and 
the exact supersymmetry plays a key role to restore the full 
supersymmetry in the continuum limit without any fine tuning of parameters. 
One of the examples is ${\cal N}=2$ Wess-Zumino model in two-dimensions 
where the Nicolai mapping was heavily utilized to construct the 
lattice model \cite{kikukawa-nakayama}. 
Also for SYM theories with extended supersymmetry, 
Cohen, Kaplan, Katz and \"{U}nsal recently proposed 
such a kind of various lattice models motivated by the idea of 
deconstruction \cite{kaplan, kaplan2}\footnote{
For some related works, see refs. \cite{kaplan_related}. 
Also, for other attempts to lattice formulations of supersymmetry, 
see \cite{other, catterall}.}.  
In these models, to generate the kinetic terms of the target theories 
and to stabilize noncompact bosonic zero-modes (the so-called radions), 
one has to add terms softly breaking the exact supersymmetry, 
which are tuned to vanish in large volume limit.   

In many cases, the above `internal' supersymmetries  
can be reinterpreted as the BRST symmetries in topological field theories. 
So, in order to construct lattice models respecting the `internal' 
supersymmetry, it seems natural to start with the topological field theory 
formulation of the theory. In ref. \cite{catterall}, 
Catterall discussed on a general formulation 
of lattice models based on the connection to topological field theory 
for supersymmetric theories without gauge symmetry.  
Two-dimensional ${\cal N}=2$ Wess-Zumino model was also investigated there. 

Here, standing on the same philosophy, 
we construct lattice models for SYM theories with extended 
supersymmetry keeping one or two supercharges exactly. 
Our models are motivated by the topological field theory formulation of 
${\cal N}=2, 4$ SYM theories, 
and free from the radion problems. 
The lattices have hypercubic structures, 
and the gauge fields are expressed as ordinary 
compact unitary variables on the lattice links. 
Our lattice models have a huge degeneracy of the classical vacua 
growing as exponential of the number of plaquettes, 
which causes difficulty for perturbative analysis around a specific vacuum. 
{}From the reason not completely same but somewhat 
similar to the case \cite{kaplan, kaplan2},  
we add terms breaking the exact supersymmetry to pick up a single vacuum 
corresponding to the desired continuum theory. 

The classical actions of ${\cal N}=2$ SYM theories 
are written as the `topological field theory (TFT) 
form': $Q$(something). $Q$ is one of the supercharges of the SYM theories and 
nilpotent up to gauge transformations.  
Topological field theories are related to the ${\cal N}=2$ theories 
via the procedure of twist \cite{witten}. 
The twist changes the spin quantum numbers of fields, 
and some of supercharges become scalar.    
When regarding one of the scalar supercharges $Q$ as the BRST charge  
whose cohomology defines physical observables, 
the theories become topological field theories.  
In this paper, 
however we do not change the interpretation of the supercharge $Q$. 
Namely, $Q$ is regarded as just one of spinor supercharges of the original SYM 
theories not as the BRST operators. 
For theories on the flat space, the twisted 
theories are equivalent to the original theories, and 
the twist is solely renaming the field variables. 
We construct the lattice theories for ${\cal N}=2$ SYM theories in the 
dimensions $d=2$, 4, 8 keeping the exact supercharge $Q$. 
The lattice actions have the $Q$-exact form as well as the continuum ones. 
Among them, the $d=2$ case is shown to flow 
to the desired continuum theory without 
any fine tuning of parameters, where the supersymmetry breaking term  
does not alter the result of renormalization.  

For ${\cal N}=4$ SYM theories, the actions are written as the form:  
$Q_+Q_-$(something), where $Q_{\pm}$ are two of the supercharges and 
nilpotent up to gauge transformations.  
We refer it as `balanced topological field theory (BTFT) form'. 
Balanced topological field theories are also related to ${\cal N}=4$ 
SYM theories via the twist \cite{vafa-witten, dijkgraaf-moore, lab}. 
Similarly to the ${\cal N}=2$ case, we regard $Q_{\pm}$ as two of spinor  
supercharges of the ${\cal N}=4$ theories, and the twist amounts just to 
changing the notations of fields. 
The lattice models are constructed for ${\cal N}=4$ SYM in two- and 
four-dimensions preserving the $Q_{\pm}$ supersymmetries exactly. 
In this case, the lattice actions have SU$(2)_R$ symmetry transforming 
$Q_{\pm}$ as a doublet, which is a subgroup of SU(4) internal symmetry 
of ${\cal N}=4$ theories. 
We show that the lattice theory for ${\cal N}=4$ in two-dimensions 
defines the corresponding continuum theory with no requirement for fine tuning 
as well as 
${\cal N}=8$ theory in two-dimensions which is obtained by 
dimensional reduction from the ${\cal N}=4$ in four-dimensions. 

This paper is organized as follows. 
In section \ref{sec:TFT}, we briefly 
explain on the `topological field theory form' 
of ${\cal N}=2$ SYM theories. In section \ref{sec:lat_N=2}, we construct 
the lattice models for ${\cal N}=2$ theories in $d=2, 4, 8$, and show that 
the continuum theory is defined without fine tuning in the case $d=2$. 
In section \ref{sec:BTFT}, we give some explanations on 
the `balanced topological field theory form' for ${\cal N}=4$ SYM theories in 
two- and four-dimensions. 
In section \ref{sec:lat_N=4}, we formulate the ${\cal N}=4$ theories on 
hypercubic lattices, and discuss on the renormalization in the continuum 
limit. Section \ref{sec:summary} is devoted to summary and discussions. 
In appendix \ref{sec:gamma_N=2}, we give the explicit form of the 
$\gamma$-matrix conventions used in the paper. 
In appendix \ref{sec:Pmu}, we present 
the explicit form of the matrices $P_{\mu}$ for the ${\cal N}=2$ 
and 4 theories, 
which appear in the term analogous to the Wilson term 
removing the species doublers of 
fermions. In appendix \ref{sec:cohomology}, we discuss 
about $Q_{\pm}$-cohomology in the ${\cal N}=4$ lattice models, 
whose result is useful for the 
renormalization argument in section \ref{sec:lat_N=4}.    
   
Throughout this paper, 
we focus on the gauge group $G = {\rm SU}(N)$. 
In sections \ref{sec:TFT} and \ref{sec:BTFT} discussing continuum 
theories, notations of repeated indices in formulas are assumed to be summed. 
On the other hand, 
in sections \ref{sec:lat_N=2} and \ref{sec:lat_N=4} for 
lattice theories, we explicitly write the summation over the indices 
except the cases of no possible confusion.  
  
\section{`Topological Field Theory Form' of ${\cal N}= 2$ Super 
Yang-Mills Theories}
\label{sec:TFT}
First, we give some explanation on 
Wick-rotated ${\cal N}=2$ SYM theories \cite{waldron}, 
and then move on their description 
in the `topological field theory form'. 

\subsection{Note on Wick-rotation and U$(1)_R$ Symmetry} 

We start with ${\cal N}=1$ SYM in $(D-1, 1)$-dimensions ($D=4,6,10$): 
\beq
S^{(M)} = \frac{1}{g^2} \int \dd^Dx\, \tr \left[-\frac14 F^{MN}F_{MN} 
-\frac{i}{2}\bar{\Psi}\Gamma^MD_M \Psi\right], 
\eeq
where $M, N= 0,1,\cdots D-1$, and the metric tensor is 
$\eta^{MN} = \diag (-1,1,1,\cdots, 1)$. 
The $\Gamma$-matrices satisfy the Clifford algebra 
$\{\Gamma^M, \, \Gamma^N\} = -2\eta^{MN}$. 
Here we take the Majorana representation for $\Gamma^M$. 
Namely, all of $\Gamma^M$ consist of pure imaginary entries.  
$\Gamma^0$ is antisymmetric, 
while $\Gamma^i$ ($i= 1, \cdots, D-1$) are symmetric. 
The spinor field is expanded as $\Psi = \sum_a \Psi^a T^a$. 
$\Psi^a$ are real Grassmann valued spinors, and 
$T^a$ represent a basis of the gauge group generators. 
The number of spinor components is $2(D-2)$. 
Note 
\beq
\bar{\Psi} = \Psi^T \Gamma^0, 
\label{bar}
\eeq
where the transpose acts only on spinor indices. 
The action is invariant under the Lorentz boost along the $x^i$ direction 
transforming the fermion as  
\beq
\Psi \rightarrow e^{\Gamma^0\Gamma^i\frac{\theta}{2}}\Psi. 
\label{boost}
\eeq

After the Wick-rotation ($x^0 \rightarrow -ix_D$, 
$\Gamma_D \equiv i\Gamma^0$), 
the Euclidean action reads 
\beq
S^{(E)} = \frac{1}{g^2} \int \dd^Dx\, \tr \left[\frac14 F_{MN}F_{MN} 
+\frac{i}{2}\bar{\Psi}\Gamma_MD_M \Psi\right], 
\label{euclid_S}
\eeq
with $\{ \Gamma_M, \Gamma_N\} = -2\delta_{MN}$ ($M,N = 1, \cdots, D$). 
Now the Lorentz boost (\ref{boost}) becomes a rotation in $(i, D)$-plane: 
\beq
\Psi \rightarrow e^{\Gamma_D\Gamma_i\frac{\theta}{2}}\Psi. 
\label{rotation}
\eeq
This rotation does not keep the property of $\Psi^a$ being real. 
Nevertheless, we can give a justification to regard the rotation as a 
symmetry of the action by still using (\ref{bar}) for $\bar{\Psi}$ 
in the action (\ref{euclid_S}).  
We can interpret $\Psi^a$ as complexified Grassmann 
and the action as holomorphic 
with respect to $\Psi^a$ without changing the result of the path 
integral\footnote{A complexified Grassmann number takes the form: 
$(\mbox{complex number})\times (\mbox{real Grassmann number})$.}.  
In the complexified space, (\ref{rotation}) is closed and the rotation can 
be understood as a symmetry of the action. 

The action (\ref{euclid_S}) is invariant under the supersymmetry 
transformation: 
\beq
\delta A_M = \epsilon^T\gamma_M \Psi, \quad
\delta \Psi = -\frac12 (F_{ij}\gamma_{ij} + 2F_{iD}\gamma_i)\epsilon, 
\label{susy}
\eeq
where $\epsilon$ is a real spinor parameter. Also,  
$\gamma_i\equiv -i\Gamma^0\Gamma_i$, 
$\gamma_D \equiv {\bf 1}$, $\gamma_{ij} = \frac12 [\gamma_i, \, \gamma_j]$. 
Then, $\gamma_i$ satisfy 
$\{ \gamma_i, \, \gamma_j \} = -2\delta_{ij}$.   

As a result of dimensional reduction with respect to the $x^{D-1}$ and 
$x^D$ directions, the theory (\ref{euclid_S}) becomes ${\cal N}=2$ SYM 
theory in $d\equiv (D-2)$-dimensions. 
The rotational symmetry in $(D-1, D)$-plane is seen as U$(1)_R$ 
internal symmetry in the ${\cal N}=2$ theory. 
Also, the ${\cal N}=2$ supersymmetry is given by the dimensional reduction 
of (\ref{susy}). 

\subsection{`Topological Field Theory Form'}

The result of the dimensional reduction of the action (\ref{euclid_S}) 
can be written as the 
`topological field theory (TFT) form' \cite{witten, brooks, kanno}: 
\beq
S_{{\cal N}=2} = Q \frac{1}{2g_d^2}\int \dd^dx \, \tr \left[
\frac14 \eta\, [\phi, \,\bar{\phi}] -i\vec{\chi}\cdot\vec{\Phi} 
+\vec{\chi}\cdot\vec{H} -i\psi_{\mu}D_{\mu}\bar{\phi}\right], 
\label{TFT_S}
\eeq
where $\mu$ is the index for $d$-dimensional space-time. 
Bosonic fields are $A_{\mu}$, $\phi \equiv A_{D-1} + i A_D$, 
$\bar{\phi} \equiv A_{D-1}-iA_D$, and the auxiliary fields $\vec{H}$. 
The other fields $\psi_{\mu}$, $\vec{\chi}$, $\eta$ are fermionic. 
$\vec{H}$ and $\vec{\chi}$ are column vectors with $(d-1)$-components.  
$\vec{\Phi}$ are functions of field strengths $F_{\mu\nu}$ and 
have one, three, seven components for the cases $d=2,4,8$, 
respectively. Their forms are given by  
\bea
 & & \Phi \equiv 2F_{12}  \quad \mbox{for } d=2 \\
 & & \Phi_{\bA} \equiv 2\left(F_{{\bA} 4} + \frac12\,\varepsilon_{\bA\bB\bC}\,
F_{\bB\bC}\right) \quad \mbox{for } d=4 \\
 & & \Phi_{\bA} \equiv 2\left(F_{{\bA} 8} + \frac12\,c_{\bA\bB\bC}\,F_{\bB\bC}
\right) \quad \mbox{for } d=8. 
\eea
$c_{\bA\bB\bC}$ are totally antisymmetric and 
expressed as the structure constants 
of the algebra of octonions. Nonzero components are 
\beq
c_{127}=c_{163}=c_{154}=c_{253}=c_{246}=c_{347}=c_{567}=1 
\eeq
and those obtained by index permutations therefrom. 

$Q$ is one of the supercharges for the transformation (\ref{susy}), 
and its transformation rule is given as   
\bea
QA_{\mu} = \psi_{\mu}, & & Q\psi_{\mu} = iD_{\mu}\phi, \nn \\
Q\phi = 0, & &    \nn \\
Q\vec{\chi} = \vec{H}, & & Q\vec{H} = [\phi, \,\vec{\chi}], \nn \\
Q\bar{\phi} = \eta, & & Q\eta = [\phi, \,\bar{\phi}]. 
\label{Q_continuum}
\eea
$Q$ is nilpotent up to infinitesimal gauge transformations 
with the parameter $\phi$. 
By choosing the representation of $\gamma_i$ so that $\Psi$ takes the form 
\beq
\Psi^T = \left( \psi_1, \cdots, \psi_d, \chi_1, \cdots, \chi_{d-1}, 
\frac12\eta \right), 
\label{Psi_N=2}
\eeq
$Q$ is expressed as the transformation (\ref{susy}) with 
$\epsilon^T = (0, \cdots, 0, -\varepsilon)$. 
$\varepsilon$ is a real Grassmann number. 
See appendix \ref{sec:gamma_N=2} for the explicit form of $\gamma_i$. 
Since in the path integral 
we may regard $\phi$ and $\bar{\phi}$ as independent variables, 
it is not necessary to 
worry about the asymmetry with respect to 
$\phi$ and $\bar{\phi}$ in the transformation (\ref{Q_continuum}). 

The action (\ref{TFT_S}) is written more explicitly as 
\bea
S_{{\cal N}=2} & = & \frac{1}{2g_d^2}\int \dd^dx\, \tr \left[
\frac14 [\phi, \,\bar{\phi}]^2 + \vec{H}\cdot\vec{H}
-i\vec{H}\cdot\vec{\Phi}+D_{\mu}\phi D_{\mu}\bar{\phi} \right. \nn \\
 & & \hspace{2cm} \left. -\frac14 \eta\,[\phi, \, \eta]
-\vec{\chi}\cdot [\phi, \, \vec{\chi}] +\psi_{\mu} [\bar{\phi}, \psi_{\mu}]
 +i\vec{\chi}\cdot Q\vec{\Phi} +i\psi_{\mu} D_{\mu}\eta
\right]. 
\label{TFT_S2}
\eea
The charge for the U$(1)_R$ rotation is assigned as follows: $+2$ for $\phi$, 
$-2$ for $\bar{\phi}$, $+1$ for $\psi_{\mu}$, 
$-1$ for $\vec{\chi}$ and $\eta$, $0$ for $A_{\mu}$ and $\vec{H}$. 
As is seen from the Yukawa interactions in (\ref{TFT_S2}), the U$(1)_R$ is a 
chiral rotation which suffers anomaly generically 
in cases\footnote{In two-dimensions, 
since 
the Pontryagin class is $\tr F$, it vanishes for 
$G={\rm SU}(N)$ and 
the chiral symmetry U$(1)_R$ is not anomalous.
} $d=4, 8$. 

In refs. \cite{witten}, topological field theory was obtained 
via the procedure of twist from 
theory with ${\cal N}=2$ supersymmetry. The twist changes the spin of 
fields and some supercharges become scalar. 
If one of the scalar supercharges is regarded as the BRST charge, whose 
cohomology and the gauge invariance determine physical observables,  
the theory becomes topological field theory. 
In the context of topological field theory, the U$(1)_R$ charge stands for 
the ghost number, and the U$(1)_R$ anomaly is seen 
as the formal dimension of instanton moduli space. The instanton is defined 
by configurations satisfying 
the equations $\vec{\Phi}=0$. Through the index theorem, 
the difference of the number of $\psi_{\mu}$ zero-modes 
(having the U$(1)_R$ charge $+1$) 
and the number of $\vec{\chi}$, $\eta$ zero-modes (having the charge $-1$) 
gives the 
formal dimension of the instanton moduli space.  

Note that in this paper 
we regard the theory (\ref{TFT_S}) as ${\cal N}=2$ SYM 
theory itself, not as topological field theory. 
In fact, (\ref{TFT_S}) is nothing but the ${\cal N}=2$ SYM action 
after renaming the fermionic fields as (\ref{Psi_N=2})\footnote{Note that 
the index $\mu$ of $\psi_{\mu}$ does not reflect the Lorentz transformation 
property.} 
up to total derivative terms. 
Here, we consider $Q$ as one of the spinor supercharges 
and do not use it to define physical observables. 
Physical operators are determined solely 
by gauge invariance of the ${\cal N}=2$ SYM theory. 
{}From this view-point, the topological field theory is considered as 
a special subsector of the original ${\cal N}=2$ supersymmetric theory. 

The total derivative terms exist in the `TFT form' (\ref{TFT_S}) for 
the cases $d=4,8$. They are explicitly written as   
\bea
 & & \frac{1}{2g_4^2}\int \, \tr (F\wedge F) \quad \mbox{for } d=4,  \nn \\
 & & \frac{1}{2g_8^2}\int \, \Omega\wedge\tr (F\wedge F) 
           \quad \mbox{for } d=8, 
\label{total_derivative}
\eea
where $F= \frac12 F_{\mu\nu}\dd x^{\mu}\wedge \dd x^{\nu}$, 
and $\Omega$ is a self-dual closed four-form: 
\bea 
\Omega & = & \omega + *\omega \nn \\
\omega & = & \frac{1}{3!}\,c_{\bA\bB\bC}\, 
\dd x^{\bA}\wedge\dd x^{\bB}\wedge\dd x^{\bC}\wedge\dd x^8.  
\eea

\setcounter{equation}{0}
\section{Lattice Formulation with One Exact Supercharge}
\label{sec:lat_N=2}

In this section, we formulate the theory (\ref{TFT_S}) 
on the $d$-dimensional hypercubic lattice 
keeping the supersymmetry $Q$. 
In the lattice theory, gauge fields $A_{\mu}(x)$ 
are promoted to the compact unitary variables 
\beq
U_{\mu}(x)= e^{iaA_{\mu}(x)}
\label{unitary}
\eeq 
on the link $(x, x+\hat{\mu})$. 
`$a$' stands for the lattice spacing, and $x\in {\bf Z}^d$ the lattice site. 
All other variables are distributed at sites.  
Interestingly, the $Q$-transformation (\ref{Q_continuum}) is extendible 
to the lattice variables preserving the property 
\beq
Q^2 = (\mbox{infinitesimal gauge transformation with the parameter } \phi)
\label{Q_nilpotent}
\eeq
as follows: 
\bea
 & & QU_{\mu}(x) = i\psi_{\mu}(x) U_{\mu}(x), \nn \\
 & & Q\psi_{\mu}(x) = i\psi_{\mu}(x)\psi_{\mu}(x) 
    -i\left(\phi(x) - U_{\mu}(x)\phi(x+\hat{\mu})U_{\mu}(x)^{\dagger}\right),
  \nn \\
 & & Q\phi(x) = 0,     \nn \\
 & & Q\vec{\chi}(x) = \vec{H}(x), \quad 
           Q\vec{H}(x) = [\phi(x), \,\vec{\chi}(x)], \nn \\
 & & Q\bar{\phi}(x) = \eta(x), \quad  Q\eta(x) = [\phi(x), \,\bar{\phi}(x)]. 
\label{Q_lattice}
\eea
Also, 
$QU_{\mu}(x)^\dagger = -iU_{\mu}(x)^\dagger \psi_{\mu}(x)$ follows from 
$U_{\mu}(x)U_{\mu}(x)^\dagger =1$. 
All transformations except $QU_{\mu}(x)$ and 
$Q\psi_{\mu}(x)$ are of the same form as in the continuum case. 
Since (\ref{Q_nilpotent}) means  
\bea
Q^2U_{\mu}(x) & = & i(Q\psi_{\mu}(x))\,U_{\mu}(x)
-i\psi_{\mu}(x)\,(QU_{\mu}(x)) 
\nn \\
 & = & \phi(x)U_{\mu}(x)-U_{\mu}(x)\phi(x+\hat{\mu}),    
\eea
if we assume the formula ``$QU_{\mu}(x) = \cdots$'',
the transformation $Q\psi_{\mu}(x)$ is determined\footnote{The first term 
in the RHS does not vanish because 
$i\psi_{\mu}(x)\psi_{\mu}(x)= 
-\frac{1}{2}f^{abc}\psi_{\mu}^a(x)\psi_{\mu}^b(x)T^c$ with $f^{abc}$ being 
structure constants of the gauge group.}. 
Then, happily  
$Q^2\psi_{\mu}(x)=[\phi(x), \,\psi_{\mu}(x)]$ is satisfied, 
and the $Q$-transformation is consistently closed. 
Note that we use the dimensionless variables here, and that 
various quantities are of the following orders:  
\bea
& & \psi_{\mu}(x), \vec{\chi}(x), \eta(x) = O(a^{3/2}), \quad 
\phi(x), \bar{\phi}(x) = O(a), \quad \vec{H}(x) = O(a^2), \nn \\
 & & Q=O(a^{1/2}). 
\label{order_of_a}
\eea
The first term in the RHS of ``$Q\psi_{\mu}(x)=\cdots$'' in (\ref{Q_lattice}) 
is of 
subleading order $O(a^3)$. 

\subsection{Lattice Action}

Once we have the $Q$-transformation rule closed among lattice variables, 
it is almost straightforward to construct the lattice action with the exact 
supersymmetry $Q$: 
\bea
S^{{\rm LAT}}_{{\cal N}=2} & = & Q\frac{1}{2g_0^2}\sum_x \, \tr\left[ 
\frac14 \eta(x)\, [\phi(x), \,\bar{\phi}(x)] -i\vec{\chi}(x)\cdot\vec{\Phi}(x) 
+\vec{\chi}(x)\cdot\vec{H}(x)\right. \nn \\
 & & \hspace{2cm}\left. \frac{}{} 
+i\sum_{\mu=1}^d\psi_{\mu}(x)\left(\bar{\phi}(x) - 
U_{\mu}(x)\bar{\phi}(x+\hat{\mu})U_{\mu}(x)^{\dagger}\right)\right], 
\label{lat_N=2_S}
\eea
where 
\bea
\Phi(x) & = & -i\left[U_{12}(x)- U_{21}(x)\right] \quad \mbox{for  } d=2, 
\label{Phi_2d} \\
\Phi_{\bA}(x) & = & -i\left[U_{{\bA} 4}(x) -U_{4\bA}(x) 
+ \frac12\sum_{{\bB},{\bC}=1}^3\varepsilon_{\bA\bB\bC}\,
(U_{\bB\bC}(x) - U_{\bC\bB}(x))\right] \quad \mbox{for  }d=4, 
\label{Phi_4d} \\
\Phi_{\bA}(x) & = & -i\left[U_{{\bA} 8}(x) -U_{8\bA}(x) 
+ \frac12 \sum_{{\bB},{\bC}=1}^7 c_{\bA\bB\bC}\,
(U_{\bB\bC}(x) - U_{\bC\bB}(x))\right] \quad \mbox{for  }d=8. 
\eea
$U_{\mu\nu}(x)$ are plaquette variables written as
\beq
U_{\mu\nu}(x) \equiv U_{\mu}(x) U_{\nu}(x+\hat{\mu}) 
U_{\mu}(x+\hat{\nu})^{\dagger} U_{\nu}(x)^{\dagger}. 
\eeq
The action (\ref{lat_N=2_S}) is clearly $Q$-invariant 
from its $Q$-exact form. 
After acting $Q$ in the RHS, the action takes the form 
\bea
S^{{\rm LAT}}_{{\cal N}=2} & = & \frac{1}{2g_0^2}\sum_x \, \tr\left[
\frac14 [\phi(x), \,\bar{\phi}(x)]^2 + \vec{H}(x)\cdot\vec{H}(x) 
-i\vec{H}(x)\cdot\vec{\Phi}(x) \right. \nn \\
 & & \hspace{1.5cm}
+\sum_{\mu=1}^d\left(\phi(x)-U_{\mu}(x)\phi(x+\hat{\mu})U_{\mu}(x)^{\dagger}
\right)\left(\bar{\phi}(x)-U_{\mu}(x)\bar{\phi}(x+\hat{\mu})
U_{\mu}(x)^{\dagger}\right) \nn \\
 & & \hspace{1.5cm} -\frac14 \eta(x)[\phi(x), \,\eta(x)] 
- \vec{\chi}(x)\cdot [\phi(x), \,\vec{\chi}(x)] \nn \\
 & & \hspace{1.5cm}
-\sum_{\mu=1}^d\psi_{\mu}(x)\psi_{\mu}(x)\left(\bar{\phi}(x)  + 
U_{\mu}(x)\bar{\phi}(x+\hat{\mu})U_{\mu}(x)^{\dagger}\right) \nn \\
 & & \hspace{1.5cm}\left. \frac{}{}+ i\vec{\chi}(x)\cdot Q\vec{\Phi}(x) 
-i\sum_{\mu=1}^d\psi_{\mu}(x)\left(\eta(x)-
U_{\mu}(x)\eta(x+\hat{\mu})U_{\mu}(x)^{\dagger}\right)\right]. 
\label{lat_N=2_S2}
\eea
After integrating out $\vec{H}(x)$, induced $\vec{\Phi}(x)^2$ term   
yields the gauge kinetic term as the form 
\beq
\frac{1}{2g_0^2}\sum_x
\sum_{\mu < \nu}\tr\left[-(U_{\mu\nu}(x) - U_{\nu\mu}(x))^2\right]
\label{gauge_kin}
\eeq
which leads double winding plaquette terms. 
Note difference from the standard Wilson action 
\beq
\frac{1}{2g_0^2}\sum_x\sum_{\mu < \nu}\tr\left[2-U_{\mu\nu}(x)-U_{\nu\mu}(x)
\right]. 
\label{standard_wilson}
\eeq
Also, for $d=4, 8$, there appear terms corresponding to the total derivatives 
(\ref{total_derivative}). 

In contrast with (\ref{standard_wilson}) giving 
the unique minimum $U_{\mu\nu}(x)=1$, 
the action (\ref{gauge_kin}) has many classical vacua  
\beq
U_{\mu\nu}(x) = \left( \begin{array}{ccc} \pm 1 &        &       \\   
                                                & \ddots &       \\
                                                &        & \pm 1
\end{array}\right) 
\label{huge_minima}
\eeq
up to gauge transformations, where any combinations of $\pm 1$ 
with `$-1$' appearing even times are allowed in the diagonal entries. 
The number of `$-1$' means strength of a color flux through the plaquette. 
Since the configurations (\ref{huge_minima}) can be 
taken freely for each plaquette, it leads a huge degeneracy of  
vacua with the number growing as exponential of the number of plaquettes. 
In order to see the dynamics of the model, we need to sum up contributions 
from all the minima,  
and the ordinary weak field 
expansion around a single vacuum 
$U_{\mu\nu}(x)=1$ can not be justified\footnote{ 
This is the same difficulty already encountered in ref. \cite{elitzur}. 
We thank Y.~Shamir for the crucial comment.}. 
Thus, we can not say anything on the continuum limit of the lattice 
model (\ref{lat_N=2_S2}) without its nonperturbative investigations.  

Aiming to resolve the degeneracy, we add the following term to the action: 
\beq
\Delta S = \frac{1}{2g_0^2}\, \rho \sum_x\sum_{\mu < \nu}\tr 
\left(2 - U_{\mu\nu}(x) - U_{\nu\mu}(x)\right)
\label{SUSY_breaking}
\eeq
which is proportional to the standard Wilson action (\ref{standard_wilson}), 
resolving the degeneracy with the split $\frac{4\rho}{g_0^2}$. 
We will take the coefficient $\rho$ 
so that $\Delta S$ vanishes in the classical 
continuum limit but then the split grows infinite. 
For example, in case of the two-dimensional $M\times M$ periodic lattice, 
the continuum limit is taken as 
$a\limit 0$ with $g_2^2 \equiv g_0^2 a^{-2}$ and $L \equiv Ma$ fixed. 
$\rho$ can be chosen as $\rho= \frac{1}{M^s}=\frac{a^s}{L^s}$ with 
the parameter $0<s<2$.  
Since the number of the lifted vacua with 
$\Delta S= \frac{4\rho}{g_0^2}= \frac{4}{g_2^2L^s}\frac{1}{a^{2-s}}$ 
is proportional to the number of plaquettes, 
the entropy effect does not obstruct the decoupling. 
While this term justifies the expansion around $U_{\mu\nu}(x)=1$, 
it breaks the supersymmetry $Q$. 
As is seen in section \ref{sec:renormalization_N=2}, however 
in the case of $d=2$ 
it does not affect the renormalization argument. 

%
%
\FIGURE{\epsfig{file=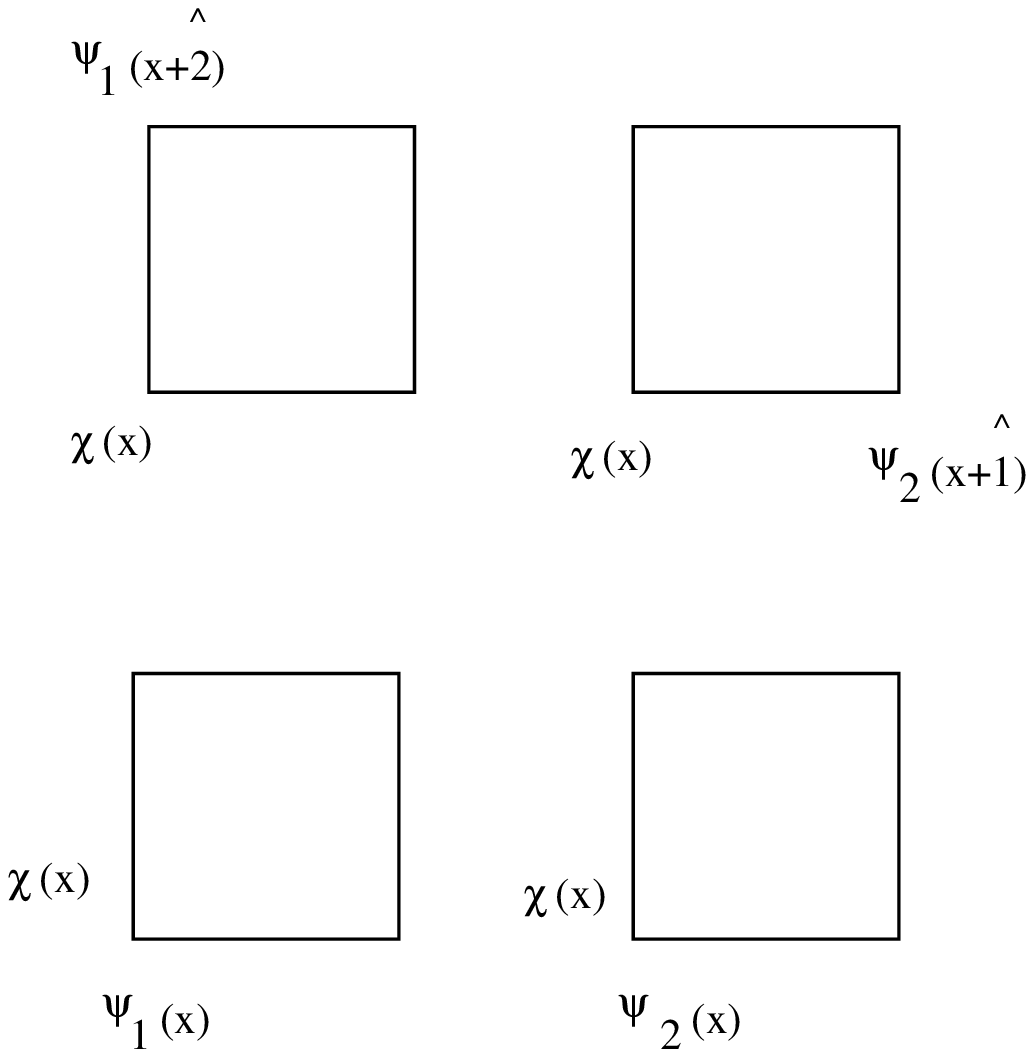,height=10cm} 
     \caption{The plaquette 
structure of the term $\vec{\chi}(x)\cdot Q\vec{\Phi}(x)$ 
for $d=2$ case. The variable $U_{\mu}$ is assigned on each link. 
Fermions $\chi$ and $\psi_{\mu}$ are distributed in the depicted four 
patterns. 
 }
     \label{fig:fermion_kin}
 }
%

The last two terms in (\ref{lat_N=2_S2}) 
give rise to the fermion kinetic term. 
In particular, $\vec{\chi}(x)\cdot Q\vec{\Phi}(x)$ has the plaquette structure 
where two fermions sit at the same site or separately at 
the nearest neighbor sites. 
As an example, Fig. \ref{fig:fermion_kin} shows the situation for 
$d=2$ case.   
Owing to this structure, the lattice model does not 
have $90$-degree rotational symmetry or the reflection positivity, either. 
However 
for $d=2$ case 
the model is shown to flow to the desired continuum theory without 
any fine tuning of parameters.

\subsection{Absence of Fermion Doubling}

We expand the exponential of the link variable (\ref{unitary}), 
and look at the kinetic terms in the action (\ref{lat_N=2_S2}).   
Because in bosonic sector no species doublers appear, 
in fermionic sector also no doublers are expected 
due to the exact supersymmetry $Q$ of (\ref{lat_N=2_S2}). 
Let us see the fermionic sector explicitly. 

After rescaling each fermion variable by $a^{3/2}$ as indicated in  
(\ref{order_of_a}), the fermion kinetic terms are expressed as 
\bea
S_f^{(2)} & = & \frac{a^4}{2g_0^2}\sum_{x, \mu}\tr\left[
-\frac12\Psi(x)^T (P_{\mu}+\gamma_{\mu})\Delta_{\mu}\Psi(x) 
+\frac12\Psi(x)^T (P_{\mu}-\gamma_{\mu})\Delta^*_{\mu}\Psi(x)\right] \nn \\
 & = & \frac{a^4}{2g_0^2}\sum_{x, \mu}\tr\left[
-\frac12\Psi(x)^T\gamma_{\mu}(\Delta_{\mu}+\Delta^*_{\mu})\Psi(x) 
-a\frac12\Psi(x)^TP_{\mu}\Delta_{\mu}\Delta^*_{\mu}\Psi(x)\right], 
\label{wilson_like_N=2}
\eea
where fermions were combined as (\ref{Psi_N=2}). 
$P_{\mu}$ are hermitian matrices satisfying 
$\{ P_{\mu}, P_{\nu}\} = 2\delta_{\mu\nu}$. 
For their explicit form, see appendix \ref{sec:Pmu}.   
$\Delta_{\mu}$ and $\Delta^*_{\mu}$ represent forward and backward difference  
operators respectively: 
\beq
\Delta_{\mu}f(x) \equiv \frac{1}{a}\left(f(x+\hat{\mu})-f(x)\right), \quad 
\Delta^*_{\mu}f(x) \equiv \frac{1}{a}\left(f(x)-f(x-\hat{\mu})\right).
\eeq
The last term containing $P_{\mu}$ has a similar structure to the Wilson 
term, and plays a role of removing fermion doublers. 
The lattice action is U$(1)_R$ invariant, and the fermion doublers are 
removed keeping the chiral U$(1)_R$. Thus, the model must break some 
assumptions of Nielsen-Ninomiya's no go theorem \cite{nielsen-ninomiya}.    
In fact, broken is 
the assumption ``{\it There exists a conserved charge $Q_F$ 
corresponding to the fermion number.}''.  
For simplicity, let us see the situation in $d=2$ case.  
When combining fermions into a two-component Dirac spinor as 
\beq
\zeta = \frac{1}{\sqrt{2}}
\left(\begin{array}{c}\psi_1 -i\psi_2 \\
                      \chi + i\frac12\eta \end{array}\right), \quad 
\bar{\zeta} = \frac{1}{\sqrt{2}}
\left(\psi_1+i\psi_2, \, \chi -i \frac12\eta\right), 
\eeq
$Q_F$ corresponds to the following U$(1)_J$ rotation: 
\beq
\zeta \rightarrow e^{i\theta}\zeta, \quad 
\bar{\zeta} \rightarrow e^{-i\theta} \bar{\zeta}. 
\eeq
The first term in (\ref{wilson_like_N=2}), 
giving a naive fermion kinetic term on the lattice, is written as  
the combination $\bar{\zeta}_{\alpha}\zeta_{\beta}$, which is invariant 
under the U$(1)_J$. On the other hand, the last term containing $P_{\mu}$ 
takes the form: 
\beq
\frac{a^4}{2g_0^2}\sum_x\frac{a}{2}\, \tr\left[ \varepsilon_{\alpha\beta} 
\zeta_{\alpha}(\Delta_1\Delta^*_1 +i\Delta_2\Delta^*_2)\zeta_{\beta} - 
\varepsilon_{\alpha\beta} 
\bar{\zeta}_{\alpha}(\Delta_1\Delta^*_1 -i\Delta_2\Delta^*_2)
\bar{\zeta}_{\beta} \right] 
\eeq
to break the U$(1)_J$ invariance.

\subsection{Renormalization} 
\label{sec:renormalization_N=2}
At the classical level, the lattice action (\ref{lat_N=2_S}) leads to 
the continuum action (\ref{TFT_S}) in the limit $a\rightarrow 0$ 
with $g_d^{-2} \equiv a^{4-d}g_0^{-2}$ kept fixed, 
and thus the ${\cal N}=2$ supersymmetry and rotational symmetry 
in $d$-dimensions are restored. 
We will check whether the symmetry restoration persists against 
quantum corrections, i.e. whether symmetry of the lattice action 
forbids any relevant or marginal operators induced which obstruct the symmetry 
restoration.  
%
\TABLE[t]{
\begin{tabular}{|c|ccccc|}
\hline \hline
$p=a+b+3c$ & 
\multicolumn{5}{|c|}{$\varphi^a\del^b\psi^{2c}$}  \\ 
\hline
0 &     &          &  1       &       &    \\
1 &     &          & $\varphi$  &      &     \\
2 &     &          & $\varphi^2$ &     &     \\
3 &     &$\varphi^3$, & $\psi\psi$, & $\varphi\del\varphi$  &     \\
4 & $\varphi^4$, & $\varphi^2\del\varphi$, & $(\del\varphi)^2$, & 
   $\psi\del\psi$, & $\varphi\psi\psi$ \\
\hline \hline
\end{tabular}
  \caption{List of operators with $p\leq 4$. 
}
\label{tab:operators}
}
%

Here, we consider the case $d=2$ only, where the U$(1)_R$ symmetry is not 
anomalous. 
%
Since the model has the critical point $g_0=0$ from the asymptotic freedom, 
we shall consider the renormalization effect perturbatively. 
For a while, we discuss without the supersymmetry breaking term $\Delta S$. 
The mass dimension of the coupling $g_2^2$ is two. 
For generic boson field $\varphi$ (other than the auxiliary fields) 
and fermion field $\psi$, the dimensions 
are 1 and 3/2 respectively.   
Thus, operators of the type $\varphi^a \del^b\psi^{2c}$ 
have the dimension $p\equiv a+b+3c$, where 
`$\del$' means a derivative with respect to the coordinates. 
{}From dimensional analysis, the operators receive the following 
radiative corrections up to some powers of possible logarithmic factors: 
\beq
\left(\frac{a^{p-4}}{g_2^2} + c_1 a^{p-2} + c_2 a^pg_2^2 + \cdots\right)
\int \dd^2x \, \varphi^a \del^b \psi^{2c},  
\label{hosei_2d}
\eeq
where $c_1, c_2, \cdots$ are constants dependent on $N$. 
The first, second and third terms in the parentheses 
represent the contributions at 
tree, one-loop 
and two-loop levels. It is easily seen from the fact that $g_2^2$ appears 
as an overall factor in front of the action and plays the same role as 
the Planck constant $\hbar$. 
Due to the super-renormalizable property in two-dimensional theory, 
the relevant corrections terminate at the two-loop. 
{}From the above formula, it is seen that 
the following operators can be relevant or marginal  
in the $a\rightarrow 0$ limit: operators with 
$p\leq 2$ induced at the one-loop level and with $p=0$ at the two-loop level. 
Operators with $p\leq 4$ are listed in Table \ref{tab:operators}. 

Since the identity operator does not affect the spectrum,    
we have to check operators of the types $\varphi$ and $\varphi^2$ only. 
Gauge symmetry and U$(1)_R$ invariance allow the operator 
$\tr \phi\bar{\phi}$, while it is forbidden by 
the supersymmetry $Q$.  
Hence,  
no relevant or marginal operators except the identity  
are generated by radiative corrections. 

When taking into account the term $\Delta S$, we still have gauge 
invariance and U$(1)_R$ symmetry but not the supersymmetry $Q$. 
This time, $\tr \phi\bar{\phi}$ might seem to remain. 
However, such is not the case. 
The supersymmetry breaking effect comes from loop diagrams 
containing the vertices of $\Delta S$, 
and insertion of $n$ $\Delta S$-vertices yields the extra factor 
$\rho^n = \frac{a^{ns}}{L^{ns}}$ to each term in (\ref{hosei_2d}). 
So we can see that contribution from the supersymmetry breaking term to 
$\tr \phi\bar{\phi}$ becomes irrelevant behaving as $a^{ns}$. 
The conclusion of the renormalization argument without $\Delta S$ is not 
affected by including the breaking term. 
In the continuum limit, 
thus full supersymmetry and rotational symmetry are 
considered to be restored without any fine tuning.

\setcounter{equation}{0}
\section{`Balanced Topological Field Theory Form' of ${\cal N}=4$ Super 
Yang-Mills Theories}
\label{sec:BTFT}

In this section, 
we consider ${\cal N}=4$ SYM theories in two- and four-dimensions, 
which are obtained by further dimensional reduction from the ${\cal N}=2$ 
theories in $d=4,\, 8$, respectively. 
The theories can be written as exact forms with respect to both of 
two supercharges, which we refer as `balanced topological field theory (BTFT) 
form' \cite{vafa-witten, dijkgraaf-moore, lab}. 
As the name `balanced' means, fields carrying the positive U$(1)_R$ charge 
are balanced with those carrying the negative charge, and thus the U$(1)_R$ 
symmetry is not anomalous. 
 
\subsection{Two-dimensional Case}

After dimensional reduction with respect to the $x^3$ and $x^4$ directions, 
the ${\cal N}=2$ SYM in $d=4$ becomes two-dimensional ${\cal N}=4$ SYM theory, 
which is written as the following `BTFT form': 
\bea
S^{2d}_{{\cal N}=4} & = &  Q_+Q_-{\cal F}^{2d}_{{\cal N}=4}, \nn \\
{\cal F}^{2d}_{{\cal N}=4} & = & \frac{1}{2g_2^2}\int  \dd^2x \, \tr\left[
-iB\Phi - \psi_{+\mu}\psi_{-\mu} - \chi_+\chi_- -\frac14\eta_+\eta_-
\right], 
\label{BTFT_S_2d}
\eea
where $Q_{\pm}$ are two of supercharges of the ${\cal N}=4$ theory 
specified below, and $\Phi \equiv 2F_{12}$. Bosons are 
gauge fields $A_{\mu}$ ($\mu=1,\,2$) and 
scalar fields $B\equiv A_3$, $C\equiv 2A_4$, $\phi$, $\bar{\phi}$. 
Also, there are auxiliary fields $\tilde{H}_{\mu}$, $H$. 
Scalars $B$ and  $C$ appear in the process of the dimensional reduction. 
Other fields $\psi_{\pm\mu}$, $\chi_{\pm}$, $\eta_{\pm}$ are 
fermions. Transformation rule of the supersymmetry $Q_{\pm}$ is given by 
\bea
& & Q_+A_{\mu}= \psi_{+\mu}, \quad Q_+\psi_{+\mu} = iD_{\mu}\phi, \quad 
Q_-\psi_{+\mu} = \frac{i}{2}D_{\mu}C-\tilde{H}_{\mu}, \nn \\
& & Q_-A_{\mu}= \psi_{-\mu}, \quad Q_-\psi_{-\mu} = -iD_{\mu}\bar{\phi}, \quad 
Q_+\psi_{-\mu} = \frac{i}{2}D_{\mu}C+\tilde{H}_{\mu}, \nn \\
 & & Q_+\tilde{H}_{\mu} = [\phi, \,\psi_{-\mu}]
-\frac12[C, \,\psi_{+\mu}] -\frac{i}{2}D_{\mu}\eta_+, \nn \\
 & & Q_-\tilde{H}_{\mu} = [\bar{\phi}, \,\psi_{+\mu}]
+\frac12[C, \,\psi_{-\mu}] +\frac{i}{2}D_{\mu}\eta_-, 
\label{group_A}
\eea
\bea
 & & Q_+B = \chi_+, \quad Q_+\chi_+ = [\phi, \,B], \quad 
Q_-\chi_+ = \frac12[C, \,B]-H, \nn \\
 & & Q_-B = \chi_-, \quad Q_-\chi_- = -[\bar{\phi}, \,B], \quad 
Q_+\chi_- = \frac12[C, \,B]+H, \nn \\
 & & Q_+H = [\phi, \,\chi_-] +\frac12[B, \,\eta_+] 
-\frac12[C, \,\chi_+],  \nn \\
 & &  Q_-H = [\bar{\phi}, \,\chi_+] -\frac12[B, \,\eta_-] 
+\frac12[C, \,\chi_-], 
\label{group_B_2d}
\eea
\bea
 & & Q_+C = \eta_+, \quad Q_+\eta_+ = [\phi, \,C], \quad 
Q_-\eta_+ = -[\phi, \,\bar{\phi}], \nn \\
 & & Q_-C = \eta_-, \quad Q_-\eta_- = -[\bar{\phi}, \,C], \quad 
Q_+\eta_- = [\phi, \,\bar{\phi}], \nn \\
 & & Q_+\phi = 0, \quad Q_-\phi= -\eta_+, \quad 
Q_+\bar{\phi} = \eta_-, \quad Q_-\bar{\phi} = 0. 
\label{group_C}
\eea
The transformation leads the following nilpotency of $Q_{\pm}$ 
(up to gauge transformations): 
\bea
Q_+^2 & = & 
(\mbox{infinitesimal gauge transformation with the parameter }\phi), \nn \\
Q_-^2 & = & 
(\mbox{infinitesimal gauge transformation with the parameter }-\bar{\phi}), 
\nn \\
\{Q_+, Q_-\} & = & 
 (\mbox{infinitesimal gauge transformation with the parameter }C). 
\label{nilpotent_N=4}
\eea
Since ${\cal F}^{2d}_{{\cal N}=4}$ is gauge invariant, 
clearly $S^{2d}_{{\cal N}=4}$ is invariant under the $Q_{\pm}$ 
transformations. Note that the form of ${\cal F}^{2d}_{{\cal N}=4}$ is not 
unique because the action is invariant under the shift 
\beq
{\cal F}^{2d}_{{\cal N}=4}\limit 
{\cal F}^{2d}_{{\cal N}=4} + Q_+G_- + Q_-G_+
\eeq
with gauge invariant $G_{\pm}$. 

With the same $\gamma$-matrix convention as the $d=4$ case in 
appendix \ref{sec:gamma_N=2}, notations for fermions in the `TFT form' 
(\ref{Psi_N=2}) correspond to   
\beq
\Psi^T = \left(\psi_{+1}, \psi_{+2}, \chi_+, \frac12\eta_+, 
\psi_{-1}, \psi_{-2}, \chi_-, \frac12\eta_-\right).   
\eeq
$Q_+$ is nothing but the dimensional reduction of $Q$ 
of the ${\cal N}=2$, $d=4$ case, and $Q_-$ represents  
the transformation (\ref{susy}) with 
$\epsilon^T = (0,0,0, -\varepsilon, 0,0,0,0)$. 
$\varepsilon$ is real Grassmann.  

Under the dimensional reduction, 
the U$(1)_R$ internal symmetry enhances to SU$(2)_R$, which is a subgroup of 
SU(4) internal symmetry of ${\cal N}=4$ theory. 
The SU$(2)_R$ generators are expressed as 
\bea
J_{++} & = & \int \dd^2x\, \left[
\psi_{+\mu}^a(x)\frac{\delta}{\delta\psi_{-\mu}^a(x)} + 
\chi_{+}^a(x)\frac{\delta}{\delta\chi_{-}^a(x)} - 
\eta_{+}^a(x)\frac{\delta}{\delta\eta_{-}^a(x)} 
+ 2\phi^a(x)\frac{\delta}{\delta C^a(x)} \right. \nn \\
 & & \hspace{1cm} \left. 
- C^a(x)\frac{\delta}{\delta \bar{\phi}^a(x)}\right],  \nn \\
J_{--} & = & \int \dd^2x\, \left[
\psi_{-\mu}^a(x)\frac{\delta}{\delta\psi_{+\mu}^a(x)} + 
\chi_{-}^a(x)\frac{\delta}{\delta\chi_{+}^a(x)} - 
\eta_{-}^a(x)\frac{\delta}{\delta\eta_{+}^a(x)} 
- 2\bar{\phi}^a(x)\frac{\delta}{\delta C^a(x)} \right. \nn \\ 
 & & \hspace{1cm} \left. 
+ C^a(x)\frac{\delta}{\delta \phi^a(x)}\right],  \nn \\
J_0 & = & \int \dd^2x\, \left[
\psi_{+\mu}^a(x)\frac{\delta}{\delta\psi_{+\mu}^a(x)} 
-\psi_{-\mu}^a(x)\frac{\delta}{\delta\psi_{-\mu}^a(x)} 
+\chi_{+}^a(x)\frac{\delta}{\delta\chi_{+}^a(x)} 
-\chi_{-}^a(x)\frac{\delta}{\delta\chi_{-}^a(x)} 
\right. \nn \\
 &  & \hspace{1cm} \left. +\eta_{+}^a(x)\frac{\delta}{\delta\eta_{+}^a(x)}
-\eta_{-}^a(x)\frac{\delta}{\delta\eta_{-}^a(x)}  
+2\phi^a(x)\frac{\delta}{\delta\phi^a(x)} 
-2\bar{\phi}^a(x)\frac{\delta}{\delta\bar{\phi}^a(x)}
\right], 
\label{SU2R}
\eea
with `$a$' being the index of a basis of the gauge group generators, 
and satisfy the algebra: 
\beq
[J_0, \, J_{++}] = 2J_{++}, \quad [J_0, \, J_{--}] = -2J_{--}, \quad 
[J_{++}, \, J_{--}] = J_0. 
\eeq
$J_0$ is a generator of the U$(1)_R$ rotation, 
which is contained in SU$(2)_R$ as its Cartan subalgebra.  
$J_{++}$ ($J_{--}$) raises (lowers) the U$(1)_R$ charge by two-units. 
For fermions, subscript $\pm$ means the U$(1)_R$ charge $\pm 1$, and 
$Q_{\pm}$ raises/lowers the charge by one. 
For auxiliary fields $\tilde{H}_{\mu}$, $H$, the charge zero is assigned. 
Under the SU$(2)_R$, each of $(\psi^a_{+\mu}, \psi^a_{-\mu})$, 
$(\chi^a_+, \chi^a_-)$, 
$(\eta^a_+, -\eta^a_-)$ and $(Q_+, Q_-)$ transforms as a doublet, 
and $(\phi^a, C^a, -\bar{\phi}^a)$ as a triplet. 

The $Q_{\pm}$-transformation properties of (\ref{group_A}), 
(\ref{group_B_2d}) and (\ref{group_C}) are symbolically summarized as 
\beq
\begin{array}{ccccc}
   &           & \psi_{+\mu} &          &   \\
   & \nearrow  &             & \searrow &      \\
A_{\mu} &      &             &          & \tilde{H}_{\mu},  \\
   & \searrow  &             & \nearrow &       \\
   &           & \psi_{-\mu} &          &       \end{array}
\hspace{2cm} 
\begin{array}{ccccc}
   &           & \chi_{+} &          &   \\
   & \nearrow  &             & \searrow &      \\
B  &           &             &          & H,     \\
   & \searrow  &             & \nearrow &       \\
   &           & \chi_{-} &          &       \end{array}
\hspace{2cm}
\begin{array}{ccc}
\phi &           &        \\
     & \searrow  &        \\
     &           & \eta_+ \\
     & \nearrow  &        \\
  C  &           &        \\
     & \searrow  &        \\
     &           & \eta_- \\
     & \nearrow  &        \\
\bar{\phi} &     &        \end{array}
\eeq
where $\nearrow$ and $\searrow$ represent the actions of $Q_+$ and $Q_-$ 
respectively. The vertical direction indicates 
the U$(1)_R$ charge (From the top, 
$+2, \, +1, \, 0, \, -1, \, -2$.). 

Finally, let us note a symmetry of the action (\ref{BTFT_S_2d}) 
under exchanging the two supercharges $Q_+ \leftrightarrow Q_-$  
with 
\bea
 & & \phi \rightarrow -\bar{\phi}, \quad \bar{\phi} \rightarrow -\phi, \quad 
B \rightarrow -B, \nn \\
 & & \chi_+ \rightarrow -\chi_-, \quad \chi_- \rightarrow -\chi_+, \quad 
\tilde{H}_{\mu} \rightarrow -\tilde{H}_{\mu}, \nn \\
 & & \psi_{\pm\mu} \rightarrow \psi_{\mp\mu}, \quad 
\eta_{\pm} \rightarrow \eta_{\mp}. 
\label{Q_exchange_2d}
\eea

\subsection{Four-dimensional Case}

Dimensional reduction from the ${\cal N}=2$ theory  
in $d=8$ with respect to the $x^5, \cdots, x^8$ directions leads 
four-dimensional ${\cal N}=4$ SYM theory.  
Its `BTFT form' is as follows: 
\bea
S^{4d}_{{\cal N}=4} & = & Q_+Q_- {\cal F}^{4d}_{{\cal N}=4}, \nn \\
{\cal F}^{4d}_{{\cal N}=4} & = & \frac{1}{2g_4^2}\int\, \dd^4x \ \tr \left[
-i\vec{B}\cdot\vec{\Phi} - 
\frac13\,\varepsilon_{\bA\bB\bC}\,B_{\bA}\,[B_{\bB}, \,B_{\bC}]
\right. \nn \\
 & & \hspace{4cm}\left. -\psi_{+\mu}\psi_{-\mu} 
- \vec{\chi}_+\cdot \vec{\chi}_- -\frac14\eta_+\eta_-\right], 
\label{BTFT_S_4d}
\eea
where $\mu=1,\cdots,4$, and $\vec{\Phi}$, $\vec{B}$, $\vec{\chi}_{\pm}$ 
are three-component vectors with the indices ${\bA, \bB, \bC} = 1,2,3$. 
$\Phi_{\bA}$ are functions of the field strengths $F_{\mu\nu}$: 
$\Phi_{\bA} = 2\left(F_{{\bA} 4} + 
\frac12\,\varepsilon_{\bA\bB\bC}\,F_{\bB\bC}\right)$, 
Bosonic field contents are gauge fields $A_{\mu}$ ($\mu=1, \cdots, 4$) and  
scalar fields $B_1\equiv -A_5$, $B_2\equiv A_6$, $B_3 \equiv A_7$, 
$C\equiv 2A_8$, $\phi$, $\bar{\phi}$. 
Also, auxiliary fields are $\tilde{H}_{\mu}$ and $H_{\bA}$. 
Scalars $\vec{B}$, $C$ arise via the dimensional reduction.
Other fields $\psi_{\pm\mu}$, $\vec{\chi}_{\pm}$ and $\eta_{\pm}$ are 
fermionic. 
The field contents are almost same as the two-dimensional ${\cal N}=4$ case 
except that each field in the quartet ($B$, $\chi_{\pm}$, $H$) is replaced 
with a three-component vector. 
For fields other than the quartet 
($\vec{B}$, $\vec{\chi}_{\pm}$, $\vec{H}$), 
the $Q_{\pm}$ transformations (\ref{group_A}) and 
(\ref{group_C}) do not change. 
The quartet transforms as 
\bea
 & & Q_+\vec{B} = \vec{\chi}_+, \quad Q_+\vec{\chi}_+ = [\phi, \,\vec{B}], 
\nn \\  
 & & Q_-\vec{B} = \vec{\chi}_-, \quad 
Q_-\vec{\chi}_- = -[\bar{\phi}, \,\vec{B}], \nn \\
 & & Q_-\chi_{+\bA}= \frac12[C, \,B_{\bA}] -\frac12\,\varepsilon_{\bA\bB\bC}\, 
[B_{\bB}, \,B_{\bC}] - H_{\bA}, \nn \\
 & & Q_+\chi_{-\bA}= \frac12[C, \,B_{\bA}] +\frac12\,\varepsilon_{\bA\bB\bC}\, 
[B_{\bB}, \,B_{\bC}] +H_{\bA}, \nn \\
 & & Q_+H_{\bA} = [\phi, \,\chi_{-\bA}] +\frac12[B_{\bA}, \,\eta_+] 
-\frac12[C, \,\chi_{+\bA}] 
-\varepsilon_{\bA\bB\bC}\, [B_{\bB}, \,\chi_{+\bC}], 
\nn \\
 & &  Q_-H_{\bA} = [\bar{\phi}, \,\chi_{+\bA}] -\frac12[B_{\bA}, \,\eta_-] 
+\frac12[C, \,\chi_{-\bA}] 
-\varepsilon_{\bA\bB\bC}\, [B_{\bB}, \,\chi_{-\bC}]. 
\label{group_B_4d}
\eea
The nilpotency (\ref{nilpotent_N=4}) is valid also here, 
and clearly seen is the $Q_{\pm}$ invariance of the action. 
With the same $\gamma$-matrix convention as in the $d=8$ case in 
appendix \ref{sec:gamma_N=2}, fermions (\ref{Psi_N=2}) correspond to  
\beq
\Psi^T = \left(\psi_{+1}, \cdots, \psi_{+4}, -\chi_{+1}, \chi_{+2}, 
\chi_{+3},\frac12\eta_+, \psi_{-1}, \cdots, \psi_{-4}, -\chi_{-1}, 
\chi_{-2}, \chi_{-3}, \frac12\eta_-\right). 
\eeq
$Q_+$ is nothing but the dimensional reduction of $Q$ 
for the ${\cal N}=2$, $d=8$ case, and $Q_-$ represents 
the transformation (\ref{susy}) with 
$\epsilon^T = (0,0,0,0,0,0,0, -\varepsilon,0,0,0,0, 0,0,0,0)$. 

Enhancement of U$(1)_R$ to SU$(2)_R$ takes place, and the generators 
have the same form as in (\ref{SU2R}) with the straightforward modifications:  
$\dd^2x \rightarrow \dd^4x$ and 
$\chi^a_{\pm}(x) \rightarrow \chi^a_{\pm {\bA}}(x)$. 
Also, the action (\ref{BTFT_S_4d}) is invariant under 
$Q_+\leftrightarrow Q_-$ with  
\bea
 & & \phi \rightarrow -\bar{\phi}, \quad \bar{\phi} \rightarrow -\phi, \quad 
\vec{B} \rightarrow -\vec{B}, \nn \\
 & & \vec{\chi}_+ \rightarrow -\vec{\chi}_-, 
\quad \vec{\chi}_- \rightarrow -\vec{\chi}_+, \quad 
\tilde{H}_{\mu} \rightarrow -\tilde{H}_{\mu}, \nn \\
 & & \psi_{\pm\mu} \rightarrow \psi_{\mp\mu}, \quad 
\eta_{\pm} \rightarrow \eta_{\mp}.  
\label{Q_exchange_4d}
\eea

\setcounter{equation}{0}
\section{Lattice Formulation with Two Exact Supercharges}
\label{sec:lat_N=4}

Similarly to the ${\cal N}=2$ cases, we consider to define the theories 
(\ref{BTFT_S_2d}), (\ref{BTFT_S_4d}) 
on hyper-cubic lattices preserving the two supercharges $Q_{\pm}$. 
Gauge fields $A_{\mu}$ are represented by the unitary variables 
$U_{\mu}(x)$ on the link $(x, x+\hat{\mu})$, while the other fields 
are put on sites.  

For both cases of two- and four-dimensions, the transformation rule 
(\ref{group_A}) is modified as 
\bea
 & & Q_+U_{\mu}(x) = i\psi_{+\mu}(x)U_{\mu}(x), \nn \\
 & & Q_-U_{\mu}(x) = i\psi_{-\mu}(x)U_{\mu}(x), \nn \\
 & & Q_+\psi_{+\mu}(x) = i\psi_{+\mu}\psi_{+\mu}(x) 
  -i\left(\phi(x)-U_{\mu}(x)\phi(x+\hat{\mu})U_{\mu}(x)^{\dagger}\right), 
\nn \\
 & & Q_-\psi_{-\mu}(x) = i\psi_{-\mu}\psi_{-\mu}(x) 
  +i\left(\bar{\phi}(x)-U_{\mu}(x)\bar{\phi}(x+\hat{\mu})U_{\mu}(x)^{\dagger}
\right), \nn \\
 & & Q_-\psi_{+\mu}(x) = \frac{i}{2}
\left\{\psi_{+\mu}(x), \,\psi_{-\mu}(x)\right\} -\frac{i}{2}
\left(C(x)-U_{\mu}(x)C(x+\hat{\mu})U_{\mu}(x)^{\dagger}\right) 
-\tilde{H}_{\mu}(x), \nn \\
 & & Q_+\psi_{-\mu}(x) = \frac{i}{2}
\left\{\psi_{+\mu}(x), \,\psi_{-\mu}(x)\right\} -\frac{i}{2}
\left(C(x)-U_{\mu}(x)C(x+\hat{\mu})U_{\mu}(x)^{\dagger}\right) 
+\tilde{H}_{\mu}(x), \nn \\
 & & Q_+\tilde{H}_{\mu}(x) = -\frac12
\left[\psi_{-\mu}(x), \,\phi(x)+U_{\mu}(x)\phi(x+\hat{\mu})U_{\mu}(x)^{\dagger}
\right] \nn \\
 & & \hspace{2cm} 
+\frac14\left[\psi_{+\mu}(x), \, C(x) +U_{\mu}(x)C(x+\hat{\mu})
U_{\mu}(x)^{\dagger}\right] \nn \\
 & & \hspace{2cm} +\frac{i}{2}\left(\eta_+(x) 
-U_{\mu}(x)\eta_+(x+\hat{\mu})U_{\mu}(x)^{\dagger}\right) \nn \\
 & & \hspace{2cm}  
+\frac{i}{2}\left[\psi_{+\mu}(x), \,\tilde{H}_{\mu}(x)\right] 
+\frac14\left[\psi_{+\mu}(x)\psi_{+\mu}(x), \,\psi_{-\mu}(x)\right], \nn \\
 & & Q_-\tilde{H}_{\mu}(x) = -\frac12
\left[\psi_{+\mu}(x), \,\bar{\phi}(x)+U_{\mu}(x)\bar{\phi}(x+\hat{\mu})
U_{\mu}(x)^{\dagger}\right] \nn \\
 & & \hspace{2cm}
-\frac14\left[\psi_{-\mu}(x), \, C(x) +U_{\mu}(x)C(x+\hat{\mu})
U_{\mu}(x)^{\dagger}\right] \nn \\
 & & \hspace{2cm} -\frac{i}{2}\left(\eta_-(x) 
-U_{\mu}(x)\eta_-(x+\hat{\mu})U_{\mu}(x)^{\dagger}\right) \nn \\
 & & \hspace{2cm}
+\frac{i}{2}\left[\psi_{-\mu}(x), \,\tilde{H}_{\mu}(x)\right] 
-\frac14\left[\psi_{-\mu}(x)\psi_{-\mu}(x), \,\psi_{+\mu}(x)\right]. 
\label{group_A_lat}
\eea
The other transformations ((\ref{group_B_2d}, \ref{group_C}) in 
two-dimensions,  (\ref{group_B_4d}, \ref{group_C}) in four-dimensions) 
do not change the form under the 
latticization. Note that this modification keeps the nilpotency 
(\ref{nilpotent_N=4}). 
  
Making use of the $Q_{\pm}$-transformation rule 
in terms of lattice variables, it is almost straightforward to 
construct lattice actions with the exact supercharges $Q_{\pm}$ 
for both of two- and four-dimensional cases: 
\bea
S^{2d\, {\rm LAT}}_{{\cal N}=4} &  = & 
Q_+Q_-\frac{1}{2g_0^2}\sum_x\, \tr \left[-iB(x)\Phi(x) - 
\sum_{\mu=1}^2\psi_{+\mu}(x)\psi_{-\mu}(x)-\chi_+(x)\chi_-(x) \right. \nn \\ 
 & & \hspace{2.5cm} \left. -\frac14\eta_+(x)\eta_-(x)\right], 
\label{lat_N=4_S_2d}\\
S^{4d\, {\rm LAT}}_{{\cal N}=4} &  = & 
Q_+Q_-\frac{1}{2g_0^2}\sum_x\, \tr \left[
-i\vec{B}(x)\cdot\vec{\Phi}(x) 
- \frac13\sum_{{\bA},{\bB},{\bC}=1}^3\varepsilon_{\bA\bB\bC}\,B_{\bA}(x)\,
[B_{\bB}(x), \,B_{\bC}(x)] \right. \nn \\ 
 & & \hspace{2.5cm}\left. 
- \sum_{\mu=1}^4\psi_{+\mu}(x)\psi_{-\mu}(x)-
\vec{\chi}_+(x)\cdot\vec{\chi}_-(x) 
-\frac14\eta_+(x)\eta_-(x)\right], 
\label{lat_N=4_S_4d}
\eea
where $\Phi(x)$ and $\vec{\Phi}(x)$ are given by 
(\ref{Phi_2d}) and (\ref{Phi_4d}), respectively. 
The orders of operators on the lattice are 
\bea
 & & \vec{B}(x), C(x), \phi(x), \bar{\phi}(x)=O(a), \quad 
\tilde{H}_{\mu}(x), \vec{H}(x) = O(a^2), \nn \\
 & & (\mbox{fermionic fields}) = O(a^{3/2}), \quad Q_{\pm} = O(a^{1/2}).
\eea

Note that the lattice formulation retains the symmetries under SU$(2)_R$ 
as well as the $Q_+\leftrightarrow Q_-$ exchange 
(\ref{Q_exchange_2d} or \ref{Q_exchange_4d}). 
Generators of SU$(2)_R$ take the same form as the continuum case 
under the trivial modification accompanied with the latticization 
($\int \dd x \rightarrow \sum_x$, etc). 
These symmetries are used later in argument for the renormalization.

\subsection{Two-dimensional Case}
Let us look closer the two-dimensional case (\ref{lat_N=4_S_2d}), and 
discuss the renormalization. 

After acting $Q_{\pm}$, the action takes the form: 
\bea
S^{2d\,{\rm LAT}}_{{\cal N}=4} & = & \frac{1}{2g_0^2}\sum_x \tr \left[
-i\left(\frac12[C(x), \,B(x)]+H(x)\right)\Phi(x) + H(x)^2 \right. \nn \\
 & & \hspace{2cm} +i\chi_-(x)Q_+\Phi(x) -i\chi_+(x)Q_-\Phi(x) 
-iB(x)Q_+Q_-\Phi(x) 
\nn \\
 & & \hspace{2cm} -[\phi(x),\,B(x)][\bar{\phi}(x), \,B(x)]
-\frac14[C(x), \,B(x)]^2 \nn \\
 & & \hspace{2cm} +\chi_+(x)[\bar{\phi}(x), \,\chi_+(x)]
-\chi_-(x)[\phi(x), \, \chi_-(x)] + \chi_-(x) [C(x), \, \chi_+(x)] \nn \\
 & & \hspace{2cm} -\chi_-(x)[B(x), \, \eta_+(x)] 
-\chi_+(x)[B(x), \, \eta_-(x)] \nn \\
 & & \hspace{2cm} -\frac14[\phi(x), \,\bar{\phi}(x)]^2 
-\frac14[\phi(x), \, C(x)][\bar{\phi}(x), \, C(x)] \nn \\
 & & \hspace{2cm} \left. -\frac14\eta_-(x)[\phi(x), \, \eta_-(x)] 
+\frac14\eta_+(x)[\bar{\phi}(x), \, \eta_+(x)] 
-\frac14\eta_+(x)[C(x), \, \eta_-(x)]\right] \nn \\
 & & +\frac{1}{2g_0^2}\sum_{x, \mu}\tr\left[
\tilde{H}_{\mu}(x)^2 
-\frac12\psi_{+\mu}(x)\psi_{+\mu}(x)\psi_{-\mu}(x)\psi_{-\mu}(x) \right. \nn \\
 & & \hspace{2cm} 
+\left(\phi(x) - U_{\mu}(x)\phi(x+\hat{\mu})U_{\mu}(x)^{\dagger}\right)
\left(\bar{\phi}(x) - U_{\mu}(x)\bar{\phi}(x+\hat{\mu})U_{\mu}(x)^{\dagger}
\right)
\nn \\
 & & \hspace{2cm}
+\frac14\left(C(x)-U_{\mu}(x)C(x+\hat{\mu})U_{\mu}(x)^{\dagger}\right)^2 
\nn \\
 & & \hspace{2cm} 
-\psi_{+\mu}(x)\psi_{+\mu}(x)\left(\bar{\phi}(x)+
U_{\mu}(x)\bar{\phi}(x+\hat{\mu})U_{\mu}(x)^{\dagger}\right) \nn \\
 & & \hspace{2cm} 
+\psi_{-\mu}(x)\psi_{-\mu}(x)\left(\phi(x)+
U_{\mu}(x)\phi(x+\hat{\mu})U_{\mu}(x)^{\dagger}\right) \nn \\
 & & \hspace{2cm}-i\psi_{+\mu}(x)\left(\eta_-(x)-
U_{\mu}(x)\eta_-(x+\hat{\mu})U_{\mu}(x)^{\dagger}\right) \nn \\
 & & \hspace{2cm}-i\psi_{-\mu}(x)\left(\eta_+(x)-
U_{\mu}(x)\eta_+(x+\hat{\mu})U_{\mu}(x)^{\dagger}\right) \nn \\
 & & \hspace{2cm}\left. 
-\frac12\left\{\psi_{+\mu}(x), \,\psi_{-\mu}(x)\right\}
\left(C(x) + U_{\mu}(x)C(x+\hat{\mu})U_{\mu}(x)^{\dagger}\right)\right]. 
\eea
Gauge kinetic terms appear from the $\Phi(x)^2$ term induced after 
$H(x)$ integrated out, whose structure is same as the ${\cal N}=2$ case. 
We add the supersymmetry breaking term (\ref{SUSY_breaking}) 
to single out the vacuum configuration $U_{\mu\nu}(x)=1$. 
$\tilde{H}_{\mu}(x)$ integrals yield kinetic terms of $B(x)$. 
Fermion kinetic terms come from the terms $\chi_{\pm}(x)Q_{\mp}\Phi(x)$ 
and $\psi_{\pm\mu}(x)(\eta_{\mp}(x)-
U_{\mu}(x)\eta_{\mp}(x+\hat{\mu})U_{\mu}(x)^{\dagger})$. 
The former has the same plaquette structure as in  
Fig. \ref{fig:fermion_kin}, and the species 
doublers are removed by the same mechanism\footnote{The fermion kinetic terms 
take the same form as (\ref{wilson_like_N=2}) with $P_{\mu}$ 
given by (\ref{Pmu_N=4}, \ref{Pmu_N=4_2d}).}.  

%
We repeat the same renormalization argument as  
in the ${\cal N}=2$, $d=2$ case. The symmetry of the 
lattice action is sufficient to restore full supersymmetry and 
rotational invariance in the continuum limit. 
For instance, 
gauge invariance and   
SU$(2)_R$ symmetry allow the operators $\tr(4\phi\bar{\phi} + C^2)$ and 
$\tr B^2$, 
but they are not admissible from the supersymmetry $Q_{\pm}$. 
For operators of the type $\varphi^2$, 
supersymmetry breaking effect by $\Delta S$ is irrelevant in two-dimensions.  
Thus, radiative corrections are not allowed to 
generate any relevant or marginal operators 
except the identity, which means that full supersymmetry and rotational 
invariance are restored in the continuum limit.

\subsection{Four-dimensional Case}

For the four-dimensional case, the plaquette structure of 
fermion kinetic terms is 
same, and fermion doublers are removed\footnote{For the explicit form of 
the matrices $P_{\mu}$, see (\ref{Pmu_N=4}, \ref{Pmu_N=4_4d}).}. 
Although it is necessary 
to add some term like (\ref{SUSY_breaking}) in order to 
resolve the vacuum degeneracy, we will not discuss it here. 
In this subsection, we shall simply see how the exact supersymmetry 
in the action (\ref{lat_N=4_S_4d}) works to 
restrict radiative corrections. 
The action (\ref{lat_N=4_S_4d}) has a symmetry under the transformation 
$x\equiv (x_1, x_2, x_3, x_4) \rightarrow 
\tilde{x}\equiv (x_3, x_1, x_2, x_4)$ 
(the cyclic permutation of $x_1$, $x_2$,  $x_3$ with $x_4$ fixed) as 
\bea
 (U_1(x), U_2(x), U_3(x), U_4(x)) &\limit &
(U_2(\tilde{x}), U_3(\tilde{x}),U_1(\tilde{x}), U_4(\tilde{x})), \nn \\
 (\psi_{\pm 1}(x), \psi_{\pm 2}(x), \psi_{\pm 3}(x), \psi_{\pm 4}(x)) 
 &\limit & (\psi_{\pm 2}(\tilde{x}), \psi_{\pm 3}(\tilde{x}), 
\psi_{\pm 1}(\tilde{x}), \psi_{\pm 4}(\tilde{x})), \nn \\
 (\tilde{H}_1(x), \tilde{H}_2(x), \tilde{H}_3(x), \tilde{H}_4(x))
  & \limit &  (\tilde{H}_2(\tilde{x}),  \tilde{H}_3(\tilde{x}), 
\tilde{H}_1(\tilde{x}), \tilde{H}_4(\tilde{x})), \nn \\
 (B_1(x), B_2(x), B_3(x))  & \limit &
(B_2(\tilde{x}), B_3(\tilde{x}), B_1(\tilde{x})), \nn \\
 (\chi_{\pm 1}(x), \chi_{\pm 2}(x), \chi_{\pm 3}(x)) & \limit &  
(\chi_{\pm 2}(\tilde{x}), \chi_{\pm 3}(\tilde{x}), \chi_{\pm 1}(\tilde{x})), 
\nn \\
 (H_1(x), H_2(x), H_3(x))  & \limit &  
(H_2(\tilde{x}), H_3(\tilde{x}), H_1(\tilde{x})), \nn \\
 (\phi(x), C(x), \bar{\phi}(x))  & \limit  & 
(\phi(\tilde{x}), C(\tilde{x}), \bar{\phi}(\tilde{x})), \nn \\ 
     \eta_{\pm}(x) & \limit & \eta_{\pm}(\tilde{x}). 
\label{perm_4fix_4d}
\eea
The cyclic permutation of $x_2$, $x_3$, $x_4$ with $x_1$ fixed: 
$x\equiv (x_1, x_2, x_3, x_4) \limit 
\tilde{x}\equiv (x_1, x_4, x_2, x_3)$ also keeps the action invariant: 
\bea
(U_1(x), U_2(x), U_3(x), U_4(x))  & \limit &  
(U_1(\tilde{x}),  U_3(\tilde{x}), U_4(\tilde{x}), U_2(\tilde{x})), \nn \\
 (\psi_{\pm 1}(x), \psi_{\pm 2}(x), \psi_{\pm 3}(x), \psi_{\pm 4}(x)) 
 & \limit &  (\psi_{\pm 1}(\tilde{x}), \psi_{\pm 3}(\tilde{x}), 
\psi_{\pm 4}(\tilde{x}), \psi_{\pm 2}(\tilde{x})), \nn \\
(\tilde{H}_1(x), \tilde{H}_2(x), \tilde{H}_3(x), \tilde{H}_4(x)) 
 & \limit &  (\tilde{H}_1(\tilde{x}),  \tilde{H}_3(\tilde{x}),
\tilde{H}_4(\tilde{x}),  \tilde{H}_2(\tilde{x})), \nn \\
 (B_1(x), B_2(x), B_3(x)) & \limit & 
 (B_3(\tilde{x}), -B_1(\tilde{x}), -B_2(\tilde{x})), \nn \\
 (\chi_{\pm 1}(x), \chi_{\pm 2}(x), \chi_{\pm 3}(x))  & \limit &  
(\chi_{\pm 3}(\tilde{x}), -\chi_{\pm 1}(\tilde{x}), 
-\chi_{\pm 2}(\tilde{x})), \nn \\
 (H_1(x), H_2(x), H_3(x))  & \limit &  
(H_3(\tilde{x}), -H_1(\tilde{x}), -H_2(\tilde{x})), \nn \\
 (\phi(x), C(x), \bar{\phi}(x))  & \limit &  
(\phi(\tilde{x}), C(\tilde{x}), \bar{\phi}(\tilde{x})), \nn \\ 
     \eta_{\pm}(x) & \limit &  \eta_{\pm}(\tilde{x}). 
\label{perm_1fix_4d}
\eea   
Thus, the action does not change under arbitrary permutations generated by 
(\ref{perm_4fix_4d}) and (\ref{perm_1fix_4d}), which contain 
$x\equiv(x_1, x_2, x_3, x_4) \limit 
\tilde{x}\equiv (x_4, x_2, x_1, x_3)$ ($x_2$: fixed) and 
$x\equiv(x_1, x_2, x_3, x_4) \limit 
\tilde{x}\equiv (x_4, x_1, x_3, x_2)$ ($x_3$: fixed). 

In analysis on the renormalization, since the mass dimension of the coupling 
$g_4^2$ is zero, we should consider 
operators with $p\leq 4$ at all loop levels. 
For $p=1,2$, there exist no possible operators due to the gauge 
invariance, SU$(2)_R$ and $Q_{\pm}$ supersymmetry. 
In $p=3$ case, it is convenient to use the result of 
$Q_{\pm}$-cohomology in appendix \ref{sec:cohomology}. 
According to the result (\ref{result_cohomology}), 
any gauge invariant and SU$(2)_R$ invariant operators $\alpha$ 
annihilated by $Q_{\pm}$ 
can be written as the form: 
\beq
\alpha = \alpha_0 + Q_+\beta_- + Q_+Q_-\gamma
\eeq
with $\alpha_0$ proportional to the identity 
operator. From dimensional analysis and the degrees $(0, 1)$ of  
$\beta_-$, the only possibility of the form of $\beta_-$ is 
$A_{\mu}\psi_{-\nu}$. Since it can not be gauge invariant, the combination 
$Q_+\beta_-$ does not exist. While the $Q_+\leftrightarrow Q_-$ symmetry 
requires 
$\gamma$ being odd ($\gamma \limit -\gamma$), 
there is no such dimension-two operator with SU$(2)_R$ 
invariance. 

In the above, we saw that symmetries of the action (\ref{lat_N=4_S_4d}) 
prevent the relevant operators with $p=1, 2, 3$ from being induced 
by loop 
corrections. For $p=4$ case, all the terms appearing in the action 
are marginal operators passing various symmetry requirements. 
The symmetries of the lattice action 
protect each structure of the following five terms 
against the quantum corrections: 
\bea
 & & Q_+Q_-\tr\left[-i\vec{B}\cdot\vec{\Phi}\right], \quad 
Q_+Q_-\tr\left[\sum_{\bA, \bB, \bC}\varepsilon_{\bA \bB \bC}\, B_{\bA}
\, [B_{\bB}, B_{\bC}]\right], \quad  
Q_+Q_-\tr\left[\sum_{\mu}\psi_{+\mu}\psi_{-\mu}\right], \nn \\
 & &  
Q_+Q_-\tr\left[\vec{\chi}_+\cdot\vec{\chi}_-\right], \quad
Q_+Q_-\tr[\eta_+\eta_-], 
\eea
while relative weights among the five possibly would receive 
some corrections.      

The subtlety on $p=4$ operators is specific in the four-dimensions. 
In the next two subsections, we will consider theories in lower dimensions 
obtained by dimensional reduction 
from the theory (\ref{lat_N=4_S_4d}), 
and see two-dimensional model flowing to the desired continuum theories 
without fine tuning.

\subsection{${\cal N}=8$ in Three-dimensions}

Dimensional reduction with respect to the $x^4$ direction from the action 
(\ref{lat_N=4_S_4d}) leads the lattice action for three-dimensional 
${\cal N}=8$ SYM. Here, the dimensional reduction simply 
means to neglect the dependence on $x^4$ for all lattice variables. 
For example, the variable $U_4(x_1, x_2, x_3, x_4)$ on the link 
$(x, x+\hat{4})$ is reduced to the site variable $V(x_1, x_2, x_3)$ 
satisfying $V(x)V(x)^{\dagger}=1$. 
The form of the action (\ref{lat_N=4_S_4d}) does not change, 
but now we should read 
\beq
U_{{\bA} 4}(x) = U_{\bA}(x) V(x+\hat{\bA})U_{\bA}(x)^{\dagger}V(x)^{\dagger}, 
\quad 
U_{4 \bA}(x) = V(x)U_{\bA}(x) V(x+\hat{\bA})^{\dagger}U_{\bA}(x)^{\dagger}. 
\eeq
Also, we add the supersymmetry breaking term 
\beq
\Delta S = \frac{1}{2g_0^2}\, \rho \left[\sum_{{\bA}<{\bB}}
\tr\left(2-U_{{\bA}{\bB}}(x)-U_{{\bB}{\bA}}(x)\right)
+\sum_{{\bA}=1}^3\tr\left(2-U_{{\bA} 4}(x)-U_{4{\bA}}(x)\right)\right], 
\eeq
where we consider on the $M\times M\times M$ periodic lattice, 
and the naive continuum limit is taken as 
$a\limit 0$ with $g_3^2\equiv g_0^2/a$ 
and $L\equiv Ma$ fixed.  
$\rho$ is chosen as  
$\rho=\frac{1}{M^s}=\frac{a^s}{L^s}$ with $0<s<1$ 
from the requirement that 
$\Delta S$ vanishes in the continuum limit 
while the levels of lifted vacua blow up.   
Under the dimensional reduction, 
symmetries of the four-dimensional model (
SU$(2)_R$, 
$Q_+ \leftrightarrow Q_-$ symmetry (\ref{Q_exchange_4d}) and 
the permutation (\ref{perm_4fix_4d})) persist in the three-dimensional 
model. 
     
Now, let us go to the renormalization argument. 
We discuss on effect of the breaking term $\Delta S$ later. 
In three-dimensions, the mass dimension of the coupling  
$g_3^2$ is one. Operators of the type $\varphi^a\del^b\psi^{2c}$ 
with the dimension $p=a+b+3c$ receive loop corrections as 
\beq
\left(\frac{a^{p-4}}{g_3^2} + c_1 a^{p-3} + c_2 a^{p-2}g_3^2 + 
c_3 a^{p-1}g_3^4 + c_4 a^pg_3^6 + \cdots\right)
\int \dd^3x\,  \varphi^a \del^b \psi^{2c},  
\label{loop_correction_3d}
\eeq 
where the notations are similar as in (\ref{hosei_2d}). From the formula, 
we read that operators with $p \leq 4-\ell$ are relevant or marginal 
at the $\ell$-loop level. 
We can use the list in Table \ref{tab:operators}. 
    
For operators with $p=1, 2$, 
gauge invariance, 
SU$(2)_R$ and $Q_+ \leftrightarrow Q_-$ symmetry 
allow 
\beq
\tr(4\phi\bar{\phi}+C^2), \quad \tr(A_4^2), \quad \tr(B_1^2+B_2^2+B_3^2), 
\quad \tr(B_1B_2+B_2B_3+B_3B_1), 
\label{varphi2_3d}
\eeq
while all of them are forbidden by 
the $Q_{\pm}$ supersymmetry.  
In order to see the $p=3$ operators, we use 
the result of $Q_{\pm}$-cohomology (\ref{result_cohomology}). 
{}From dimensional analysis, candidates for $\beta_-$ are  
$\tr (A_4\psi_{-\mu})$, but they are not $Q_-$ invariant. 
Hence there is no operator of the type $Q_+\beta_-$. 
Symmetries under $Q_+\leftrightarrow Q_-$ and the permutation 
$(x_1, x_2, x_3) \limit (x_3, x_1, x_2)$ reduce candidates for 
$\gamma$ to $\tr((B_1+B_2+B_3)A_4)$ alone. 

Loop diagrams with $n$ vertices of $\Delta S$ inserted induce 
supersymmetry breaking effect, which is suppressed by the 
factor $\frac{a^{ns}}{L^{ns}}$ for each loop contribution 
in (\ref{loop_correction_3d}). 
The power of `$a$' for the $p=2$ operators at the one-loop is $a^{ns-1}$, 
so the operators (\ref{varphi2_3d}) remain relevant for $n=1$ and 
can not be excluded. 
On the other hand, the $p=3$ operators becomes irrelevant 
when $\Delta S$-vertices inserted. 

Therefore, the radiative corrections are allowed to generate the operator 
$Q_+Q_-\tr((B_1+B_2+B_3)A_4)$ and those in (\ref{varphi2_3d}).  
In order to reach the desired supersymmetric continuum theory, 
we have to add counter terms for the operators and cancel them with 
the radiative corrections by tuning parameters. 

\subsection{${\cal N}=8$ in Two-dimensions}

As a result of further dimensional reduction with respect to $x^3$, 
we obtain a lattice theory for ${\cal N}=8$ in two-dimensions. 
In addition to the site variable $V(x)$ coming from $U_4(x)$, 
there appears a new site variable $W(x)$, which arises from $U_3(x)$, 
satisfying $W(x)W(x)^{\dagger}=1$. 
The form of the action (\ref{lat_N=4_S_4d}) is still valid with 
\bea
 & & U_{\mu 4}(x)= U_{4 \mu}(x)^{\dagger}=
U_{\mu}(x)V(x+\hat{\mu})U_{\mu}(x)^{\dagger}V(x)^{\dagger}, 
\nn \\  
 & & U_{\mu 3}(x) = U_{3 \mu}(x)^{\dagger} 
= U_{\mu}(x)W(x+\hat{\mu})U_{\mu}(x)^{\dagger}W(x)^{\dagger}, 
\nn \\
 & & U_{34}(x) = U_{43}(x)^{\dagger} = W(x)V(x)W(x)^{\dagger}V(x)^{\dagger}. 
\label{variables_2dN8}
\eea
Here, $\mu =1, 2$ for gauge variables. 
Similar to the previous cases, we consider the $M\times M$ periodic lattice  
and the supersymmetry breaking term 
\beq
\Delta S = \frac{1}{2g_0^2}\, \rho \sum_x\sum_{1\leq k < l \leq 4}
\tr\left(2-U_{kl}(x) -U_{lk}(x)\right), 
\eeq
with $\rho = \frac{1}{M^s}=\frac{a^s}{L^s}$ ($0<s<2$). 

{}From the same renormalization argument, it is seen that 
loop corrections are not allowed to 
generate any relevant or marginal operators 
except the identity, which means restoration of full supersymmetry and 
rotational invariance in the continuum limit.

\setcounter{equation}{0}
\section{Summary and Discussions}
\label{sec:summary}

In this paper, SU$(N)$ super Yang-Mills theories with extended 
supersymmetry have been constructed on hyper-cubic lattices 
of various dimensions, keeping one or two supercharges exactly. 
Our lattice models have a huge degeneracy of the classical vacua, which 
does not allow the standard argument based on the lattice perturbation 
theory. In order to avoid the difficulty,  
we added to the action some supersymmetry breaking term, 
which becomes invisible in the naive continuum limit. 
Among the lattice models, the desired continuum theories are obtained 
without any fine tuning of parameters for the cases ${\cal N}=2, 4, 8$ 
in two-dimensions. 
We have arrived at the conclusion 
checking that symmetries of the lattice action forbid 
radiative corrections to generate any relevant or marginal operators 
which prevent the lattice theory from flowing to the continuum theory. 
It is interesting to see the restoration of full supersymmetry and 
rotational invariance explicitly by computer simulation.  

To resolve the vacuum degeneracy, 
it would be also possible to impose the admissibility condition 
on each plaquette variable:  
\beq
||1-U_{\mu\nu}(x)|| < \epsilon  
\label{admissibility}
\eeq
similar to that for gauge fields coupled to 
the Ginsparg-Wilson fermions \cite{ginsparg-wilson}. 
It breaks the supersymmetry picking up contribution from the 
boundary $||1-U_{\mu\nu}(x)|| =\epsilon$, but the breaking effect 
would vanish in the weak coupling limit. 
It will be interesting to pursue the possibility and to try reformulating 
our models in terms of the Ginsparg-Wilson fermions. 

It is also worth to be investigated 
the dynamics of our lattice actions without the supersymmetry breaking terms, 
nonperturbative understanding of which will be necessary 
to identify the corresponding continuum theories.  

The lattice actions have no symmetry under discrete rotations by 90 degrees 
due to the specific couplings between fermions $\chi$ and $\psi$. 
It is possible to modify the couplings 
keeping the supersymmetry 
so that the theories are symmetric 
under the 90-degree rotations. At the same time, however doublers arise from 
both of bosonic and fermionic sectors, 
as expected from the manifest supersymmetry. 
It is interesting to identify the corresponding continuum theories  
for both cases with and without the supersymmetry breaking terms. 

While we have focussed on pure Yang-Mills theories, 
it is possible to extend our formulation to the case of matter fields coupled, 
since the actions of ${\cal N}=2, 4$ supersymmetric QCD are rewritten as 
`TFT and BTFT forms' \cite{sako}.

\acknowledgments
The author would like to thank H.~Fukaya, Y.~Kikukawa, T.~Onogi and T.~Takimi 
for valuable discussions, 
and M.~Fukuma, S.~Iso, H.~Kunitomo, M.~Ninomiya, S.~Ohta and S-J.~Rey  
for useful comments. 
He also thank the theory group members of KEK, YITP and RIKEN 
for their warm hospitality during his stay. 
Finally, he would like to thank Y.~Shamir for crucial comments on the first 
version of this paper. 


\appendix
\section{Explicit Form of $\gamma_i$ for ${\cal N}=2$}
\label{sec:gamma_N=2}
\setcounter{equation}{0}
\renewcommand{\theequation}{A.\arabic{equation}}

Here, we present the explicit form of the matrices $\gamma_i$ 
employed for `TFT form' of ${\cal N}=2$ SYM theories. 
$\gamma_i$ is written as the form 
\beq
\gamma_i =   -i\left( \begin{array}{cc} 0 & \mu_i \\
                                \mu_i^T &   0  \end{array}\right) 
\quad \mbox{for  } i=1, \cdots, D-2, \qquad
\gamma_{D-1} = i\sigma_3 \otimes {\bf 1}_{D-2},  
\eeq
where $\sigma_1$, $\sigma_2$, $\sigma_3$ denote the Pauli matrices, and 
the explicit forms of $\mu_i$ are as follows. 

For $d=2$, 
\beq
\mu_1 = \sigma_1, \quad 
\mu_2 = - \sigma_3.     
\eeq

For $d=4$, 
\beq
 \mu_1 = \left(\begin{array}{cc}
           & \sigma_1 \\ 
 -i\sigma_2 &         \end{array}\right), \quad 
\mu_2 = \left(\begin{array}{cc}
            & -\sigma_3 \\ 
 {\bf 1}_2  &          \end{array}\right), \quad
\mu_3 = \left(\begin{array}{cc}
i\sigma_2  &          \\ 
            & \sigma_1 \end{array}\right), \quad
\mu_4 = \left(\begin{array}{cc}
 -{\bf 1}_2 &           \\ 
            & -\sigma_3  \end{array}\right). 
\eeq

For $d=8$, 
\bea
 & & \mu_1 = \left(\begin{array}{cccc}
       &         &        & \sigma_1 \\   
       &         & -\sigma_1 &       \\
       & \sigma_1&        &          \\
-i\sigma_2 &    &        &           \end{array}\right), \quad            
\mu_2 = \left(\begin{array}{cccc}
       &         &        & -\sigma_3 \\   
       &         & -\sigma_3 &       \\
       & \sigma_3&        &          \\
{\bf 1}_2 &    &        &           \end{array}\right), \quad            
\mu_3 = \left(\begin{array}{cccc}
       &         & \sigma_1&         \\   
       &         &         & \sigma_1 \\
 -\sigma_1&      &        &          \\
       & -i\sigma_2 &        &           \end{array}\right), \nn \\
 & & \mu_4 = \left(\begin{array}{cccc}
       &         & \sigma_3 &        \\   
       &         &          & -\sigma_3 \\
 -\sigma_3&         &        &          \\
       & {\bf 1}_2 &        &           \end{array}\right), \quad            
\mu_5 = \left(\begin{array}{cccc}
       & -\sigma_1 &        &          \\   
 \sigma_1&         &        &       \\
       &           &        & \sigma_1 \\
       &           & -i\sigma_2  &       \end{array}\right), \quad            
\mu_6 = \left(\begin{array}{cccc}
       & -\sigma_3 &       &         \\   
 \sigma_3&         &         &       \\
         &         &        & -\sigma_3 \\
       &           & {\bf 1}_2 &         \end{array}\right), \nn \\            
 & & \mu_7 = \left(\begin{array}{cccc}
i\sigma_2&         &        &        \\   
       & i\sigma_2 &          &        \\
        &         & i\sigma_2 &          \\
       &           &        & \sigma_1  \end{array}\right), \quad            
\mu_8 = \left(\begin{array}{cccc}
-{\bf 1}_2  &       &        &          \\   
       & -{\bf 1}_2 &        &       \\
       &           & -{\bf 1}_2&        \\
       &           &         & -\sigma_3  \end{array}\right). 
\eea

The cases $d=4, 8$ correspond also to `BTFT form' for ${\cal N}=4$ SYM 
theories 
in two- and four-dimensions, respectively.

\section{Explicit Form of $P_{\mu}$ for ${\cal N}=2, 4$}
\label{sec:Pmu}
\setcounter{equation}{0}
\renewcommand{\theequation}{B.\arabic{equation}}

For both cases ${\cal N}=2$ and ${\cal N}=4$, fermion kinetic terms 
in the lattice actions are expressed as the form (\ref{wilson_like_N=2}). 
We present the explicit form of matrices $P_{\mu}$ 
appearing in the expressions. 

In the ${\cal N}=2$ theories, 
$P_{\mu}$ are written as the form  
\beq
P_{\mu} = \left(\begin{array}{cc} 0 & \nu_{\mu} \\ 
                                  \nu_{\mu} & 0    \end{array}\right)
= \sigma_1\otimes \nu_{\mu} \quad \mbox{for  }\mu = 1, \cdots,  d-1, \qquad  
P_d = \sigma_2\otimes {\bf 1}_d. 
\eeq
For each case of $d=2$, 4, 8, $\nu_{\mu}$ are given as follows. 

For $d=2$ case, 
\beq
\nu_1 = \sigma_2.
\eeq

For $d=4$ case, 
\beq
\nu_1 = \sigma_1\otimes \sigma_2, \quad \nu_2 = \sigma_2 \otimes {\bf 1}_2, 
\quad \nu_3 = -\sigma_3\otimes \sigma_2. 
\eeq

For $d=8$ case, 
\bea
 & & \nu_1 = \left(\begin{array}{cccc} 
        &         &         & \sigma_2 \\
        &         & -i\sigma_1 &       \\
        & i\sigma_1 &       &          \\
\sigma_2 &        &         &          \end{array}\right), \quad 
\nu_2 = \left(\begin{array}{cccc} 
        &         &         & -i{\bf 1}_2 \\
        &         & -i\sigma_3 &       \\
        & i\sigma_3 &       &          \\
i{\bf 1}_2 &        &         &          \end{array}\right), \quad       
\nu_3 = \left(\begin{array}{cccc} 
        &         & i\sigma_1 &        \\
        &         &           & \sigma_2 \\
-i\sigma_1 &         &       &          \\
        & \sigma_2 &         &          \end{array}\right), \nn \\
& &  \nu_4 = \left(\begin{array}{cccc} 
        &         & i\sigma_3 &        \\
        &         &           & -i{\bf 1}_2\\
-i\sigma_3 &         &       &          \\
        & i{\bf 1}_2 &         &          \end{array}\right), \quad 
\nu_5 = \left(\begin{array}{cccc} 
        & -i\sigma_1&         &          \\
i\sigma_1&         &         &       \\
        &          &         &  \sigma_2\\
        &          & \sigma_2&          \end{array}\right), \quad       
\nu_6 = \left(\begin{array}{cccc} 
        & -i\sigma_3&           &        \\
i\sigma_3&         &           &         \\
         &         &         & -i{\bf 1}_2 \\
        &          &  i{\bf 1}_2&          \end{array}\right), \nn \\
& & \nu_7 =  \left(\begin{array}{cccc} 
 -\sigma_2&          &           &        \\
         & -\sigma_2&           &         \\
         &         &  -\sigma_2&          \\
        &          &           & \sigma_2 \end{array}\right). 
\eea

In the ${\cal N}=4$ theories, $P_{\mu}$ are expressed as 
\beq
P_{\mu} = -i \left(\begin{array}{cc} 0 & \omega_{\mu} \\
                        -\omega_{\mu}^T & 0 \end{array} \right), 
\label{Pmu_N=4}
\eeq
where the explicit forms of $\omega_{\mu}$ are presented as below. 

For two-dimensional case, 
\beq
\omega_1 = \left(\begin{array}{cc}   & i\sigma_2 \\
                         -\sigma_1   &   \end{array} \right), 
\quad
\omega_2 = \left(\begin{array}{cc}   & {\bf 1}_2 \\
                          \sigma_3   &   \end{array} \right). 
\label{Pmu_N=4_2d}
\eeq

For four-dimensional case, 
\bea
 & & \omega_1 = \left(\begin{array}{cccc}
        &           &          &  i\sigma_2  \\
        &           & \sigma_1 &             \\
        & \sigma_1  &          &             \\
-\sigma_1 &         &          &             \end{array} \right), \quad 
\omega_2 = \left(\begin{array}{cccc}
        &           &          &  {\bf 1}_2  \\
        &           & \sigma_3 &             \\
        & \sigma_3  &          &             \\
\sigma_3 &         &          &             \end{array} \right), \quad 
\omega_3 = \left(\begin{array}{cccc}
        &           & -\sigma_1 &            \\
        &           &           & i\sigma_2 \\
-\sigma_1&           &          &             \\
        & -\sigma_1 &          &             \end{array} \right), \nn \\
 & & 
\omega_4 = \left(\begin{array}{cccc}
        &           & -\sigma_3 &            \\
        &           &           & {\bf 1}_2 \\
-\sigma_3&           &          &             \\
        & \sigma_3  &          &             \end{array} \right).
\label{Pmu_N=4_4d}
\eea     

\section{$Q_{\pm}$-cohomology on Lattice}
\label{sec:cohomology}
\setcounter{equation}{0}
\renewcommand{\theequation}{C.\arabic{equation}}

In this appendix, we discuss on the equivalent 
cohomology based on two supercharges 
$Q_+$, $Q_-$ nilpotent up to gauge transformations in the theory 
(\ref{lat_N=4_S_4d})\footnote{The argument is valid also for the 
two-dimensional case (\ref{lat_N=4_S_2d}) with reading the three-component 
vectors $\vec{B}$, $\vec{\chi}_{\pm}$, $\vec{H}$ as the one-component 
$B$, $\chi_{\pm}$, $H$.}. 
The basic fact is given in Theorem 2.1 in ref. \cite{dijkgraaf-moore}, 
and this gives a generalization in the context of lattice 
gauge theory. 

First, we assign the degrees $(q_+, q_-)$ to each variable as follows: 
\beq
\begin{array}{cc}
U_{\mu}: & (0,0) \\
\psi_{+\mu}: & (1,0) \\
\psi_{-\mu}: & (0, 1) \\
\tilde{H}_{\mu}: & (1,1) 
\end{array}
\qquad 
\begin{array}{cc}
\vec{B}: & (1,1) \\
\vec{\chi}_{+}: & (2,1) \\
\vec{\chi}_{-}: & (1, 2) \\
\vec{H}: & (2,2) 
\end{array}
\qquad 
\begin{array}{cc}
\phi: & (2,0) \\
\bar{\phi}: & (0,2) \\
C: & (1, 1) \\
\eta_+: & (2,1) \\
\eta_-: & (1,2) 
\end{array}
\, . 
\eeq
The charges $q_{\pm}$ are measured by the operators: 
\bea
L_+ & = & \sum_{x,a}\left[\psi^a_{+\mu}(x)
\frac{\del}{\del \psi^a_{+\mu}(x)}+\tilde{H}^a_{\mu}(x) 
\frac{\del}{\del \tilde{H}^a_{\mu}(x)} + B_{\bA}^a(x)
\frac{\del}{\del B_{\bA}^a(x)} 
+ 2\chi_{+\bA}^a(x)\frac{\del}{\del \chi_{+\bA}^a(x)} \right. \nn \\ 
 & & \hspace{1cm} +\chi_{-\bA}^a(x)\frac{\del}{\del \chi_{-\bA}^a(x)} 
+2H_{\bA}^a(x)\frac{\del}{\del H_{\bA}^a(x)} 
+ C^a(x)\frac{\del}{\del C^a(x)} + 2\phi^a(x)\frac{\del}{\del \phi^a(x)} \nn \\
 & & \hspace{1cm} \left. +2\eta_+^a(x)\frac{\del}{\del \eta_+^a(x)} + 
\eta_-^a(x)\frac{\del}{\del \eta^a_-(x)}\right], \nn \\
L_- & = & \sum_{x,a}\left[\psi^a_{-\mu}(x)
\frac{\del}{\del \psi^a_{-\mu}(x)}+\tilde{H}^a_{\mu}(x) 
\frac{\del}{\del \tilde{H}^a_{\mu}(x)} + B_{\bA}^a(x)
\frac{\del}{\del B_{\bA}^a(x)} 
+ 2\chi_{-\bA}^a(x)\frac{\del}{\del \chi_{-\bA}^a(x)} \right. \nn \\ 
 & & \hspace{1cm}  +\chi_{+\bA}^a(x)\frac{\del}{\del \chi_{+\bA}^a(x)} 
+2H_{\bA}^a(x)\frac{\del}{\del H_{\bA}^a(x)} 
+ C^a(x)\frac{\del}{\del C^a(x)} + 
2\bar{\phi}^a(x)\frac{\del}{\del \bar{\phi}^a(x)} \nn \\
 & & \hspace{1cm} \left. +2\eta_-^a(x)\frac{\del}{\del \eta_-^a(x)} + 
\eta_+^a(x)\frac{\del}{\del \eta^a_+(x)}\right], 
\eea
where `$a$' is the index with respect to the gauge group generators, and 
the summation over the indices $\mu$ and $\bA$ is assumed. 
The U$(1)_R$ charge is given by $q_+ - q_-$ and $J_0 = L_+ - L_-$. 
$Q_{\pm}$ and $J_{\pm\pm}$ shift the degrees as 
\bea
 Q_+ :\, (q_+, q_-) \limit (q_+ +1, q_-), & \quad & 
 Q_- :\, (q_+, q_-) \limit (q_+, q_-+1), \nn \\
J_{++} :\, (q_+, q_-) \limit (q_+ +1, q_--1), & \quad & 
J_{--} :\, (q_+, q_-) \limit (q_+ -1, q_-+1). 
\eea

Next, it is convenient for later argument 
to introduce the `homotopy operators' $K_{\pm}$:  
\bea
K_+ & = & -\sum_{x, a}\left[\psi^a_{+\mu}(x)
\frac{\del}{\del \tilde{H}^a_{\mu}(x)} + 2\chi_{+\bA}^a(x)
\frac{\del}{\del H_{\bA}^a(x)} - B_{\bA}^a(x)
\frac{\del}{\del \chi_{-\bA}^a(x)} 
+2\phi^a(x)\frac{\del}{\del \eta_+^a(x)} \right. \nn \\ 
 & & \hspace{1cm} \left. -C^a(x)\frac{\del}{\del \eta_-^a(x)}\right], \nn \\
K_- & = & \sum_{x, a}\left[\psi^a_{-\mu}(x)
\frac{\del}{\del \tilde{H}^a_{\mu}(x)} + 2\chi_{-\bA}^a(x)
\frac{\del}{\del H_{\bA}^a(x)} + B_{\bA}^a(x)
\frac{\del}{\del \chi_{+\bA}^a(x)} 
+2\bar{\phi}^a(x)\frac{\del}{\del \eta_-^a(x)} \right. \nn \\ 
 & & \hspace{1cm} \left. +C^a(x)\frac{\del}{\del \eta_+^a(x)}\right], 
\eea
with changing the degrees as 
\beq
K_+:\, (q_+, q_-) \limit (q_+, q_--1), \quad 
K_-:\, (q_+, q_-) \limit (q_+-1, q_-). 
\eeq
Since there is no operator with negative $q_{\pm}$, for any operator $\alpha$ 
with $(q_+, q_-)$ 
\bea
J_{++}\alpha = K_+\alpha = 0 &  \quad &  \mbox{if  }q_- = 0, \nn \\
J_{--}\alpha = K_-\alpha = 0 & \quad  & \mbox{if  }q_+ = 0. 
\eea
After some calculation, it turns out that the following algebra holds: 
\bea
\{ Q_+, \, K_+ \} = -J_{++}, & \quad & \{Q_-, \, K_+ \} = L_+, \nn \\
\{ Q_-, \, K_- \} = -J_{--}, & \quad & \{ Q_+, \, K_- \} = L_-,  
\label{QK-alg}
\eea
which plays a key role for the argument on $Q_{\pm}$-cohomology below. 

Suppose that $\alpha$ is a gauge invariant and SU$(2)_R$ invariant operator 
with the degrees $(q_+, q_-)$, and that $Q_{\pm}$ annihilate $\alpha$: 
$Q_{\pm}\alpha=0$.  
Since $K_-Q_+\alpha=0$, making use of (\ref{QK-alg}) leads 
\beq
\alpha = \left\{ \begin{array}{cl}
Q_+\left(\frac{1}{q_-}K_-\alpha\right) & \mbox{ for  }q_+\geq 1,\, q_- \geq 1, 
\\
0 & \mbox{ for  }q_+ =0, \, q_- \geq 1. \end{array} \right. 
\label{*}
\eeq
When $q_-=0$, $Q_+$-cohomology can be nontrivial. 
Similarly, starting with $K_+Q_-\alpha=0$, we arrive at 
\beq
\alpha = \left\{ \begin{array}{cl}
Q_-\left(\frac{1}{q_+}K_+\alpha\right) & \mbox{ for  }q_+\geq 1,\, q_- \geq 1, 
\\
0 & \mbox{ for  }q_+ \geq 1, \, q_- =0.  \end{array} \right. 
\label{**}
\eeq
Nontrivial $Q_-$-cohomology can appear when $q_+=0$. 

Let us consider the case $q_+ \geq 1$, $q_-\geq 1$. 
Note that $K_-\alpha$ is annihilated by $Q_-$ as a consequence of 
(\ref{QK-alg}). Thus, by the same argument to (\ref{*}), 
\beq
K_-\alpha = Q_-\left(\frac{1}{q_+-1}K_+K_-\alpha\right) \quad 
\mbox{for  }q_+\geq 2,\, q_- \geq 1. 
\eeq
Plugging this into (\ref{*}) leads 
\beq
\alpha = Q_+Q_-\left(\frac{1}{q_-(q_+-1)}K_+K_-\alpha\right)  \quad 
\mbox{for  } q_+\geq 2, \,q_-\geq 1.
\eeq
{}From the similar consideration on $K_+\alpha$, we obtain 
\beq
\alpha =  Q_-Q_+\left(\frac{1}{q_+(q_--1)}K_-K_+\alpha\right)  \quad 
\mbox{for  } q_+\geq 1, \,q_-\geq 2.
\eeq

In the case $q_+=q_-=0$, $\alpha$ is a function of $U_{\mu}$. 
But, from the condition of $Q_{\pm}$-closed, it must be proportional to 
the identity operator. 
Finally, when $q_+=q_-=1$, $\alpha$ is expressed as 
$\alpha = Q_+(K_-\alpha)= Q_-(K_+\alpha)$, where 
$K_{\pm}\alpha$ are $Q_{\pm}$-closed and expanded by $\psi_{\pm\mu}$, 
respectively. 

The above result is summarized as the following statement: \\
\noindent 
{\it Any gauge invariant and SU$(2)_R$ invariant operator $\alpha$ annihilated 
by $Q_{\pm}$ can be expressed as   
\bea
\alpha & = &  \alpha_0 + Q_+\beta_- + Q_+Q_-\gamma \nn \\
       & = &  \alpha_0 + Q_-\beta_+ + Q_+Q_-\gamma. 
\label{result_cohomology}
\eea
$\alpha_0$ is proportional to the identity, and $\beta_{\pm}$, $\gamma$ are 
gauge invariant. 
$\beta_{+}$ ($\beta_-$) is a $Q_{+}$- ($Q_-$-) closed operator of
the degrees $(1,0)$ ($(0,1)$), and has nontrivial cohomology with 
respect to $Q_{+}$ ($Q_-$).}

The argument is valid also for lattice SYM theories obtained by dimensional 
reduction from (\ref{lat_N=4_S_4d}), for instance ${\cal N}=8$ 
theories in two- and three-dimensions.


\end{document}